\documentclass{article}
\usepackage[utf8]{inputenc}
\usepackage[margin=1in]{geometry}

\usepackage[mathscr]{eucal}
\usepackage{xcolor}
\usepackage{physics}
\usepackage{amsmath}
\usepackage{algorithm}
\usepackage[noend]{algpseudocode}
\usepackage{tikz}
\usepackage{mathdots}
\usepackage{yhmath}
\usepackage{cancel}
\usepackage{color}
\usepackage{siunitx}
\usepackage{array}
\usepackage{multirow}
\usepackage{amssymb}
\usepackage{gensymb}
\usepackage{tabularx}
\usepackage{booktabs}
\usepackage{amsthm}
\usepackage{graphicx}
\usetikzlibrary{fadings}
\usetikzlibrary{patterns}
\usetikzlibrary{shadows.blur}
\usetikzlibrary{shapes}
\usepackage{stmaryrd}
\usepackage{caption}
\usepackage{subcaption}
\usepackage{wrapfig}
\usepackage{authblk}
\usepackage{hyperref}

\usepackage{multirow}

\usepackage{todonotes}

\usepackage[english]{babel}

\usepackage{amsthm}

\theoremstyle{definition}
\newtheorem{definition}{Definition}[section]

\DeclareMathOperator*{\argmin}{arg\,min}

\title{Polytopic Analysis of Music}
\author[1]{Marmoret Axel}
\author[1]{Jérémy E. Cohen}
\author[1]{Frédéric Bimbot}
\affil[1]{Univ Rennes, Inria, CNRS, IRISA, France.}
\date{2021}

\begin{document}

\maketitle

\section*{Abstract}
Structural segmentation of music refers to the task of finding a symbolic representation of the organisation of a song, reducing the musical flow to a partition of non-overlapping segments. Under this definition, the musical structure may not be unique, and may even be ambiguous. One way to resolve that ambiguity is to see this task as a compression process, and to consider the musical structure as the optimization of a given compression criteria.

In that viewpoint, C. Guichaoua~\cite{theseGuichaoua} developed a compression-driven model for retrieving the musical structure, based on the ``System and Contrast'' model~\cite{bimbot2016system}, and on polytopes, which are extension of n-hypercubes. We present this model, which we call ``polytopic analysis of music'', along with a new open-source dedicated toolbox called \textit{MusicOnPolytopes}\footnote{https://gitlab.inria.fr/amarmore/musiconpolytopes} (in Python). This model is also extended to the use of the Tonnetz as a relation system. Structural segmentation experiments are conducted on the RWC Pop dataset~\cite{goto2002rwc}. Results show improvements compared to the previous ones, presented in~\cite{theseGuichaoua}.

\section{Introduction}
Structural segmentation of music is an important task in the Music Information Retrieval (MIR) community. This task aims at representing musical information at a mesoscopic scale with symbolic information, such as letters or semantic information (verse, chorus, etc). The musical content is hence partitioned and organized into a list of segments. Relevant structural segments must be computed from low-level musical information, thus necessitating the definition of salient metrics to form and evaluate potential segments.

This work presents a compression-based scheme for structural segmentation of symbolic music (\text{i.e.} music discretized both in time and representation as a flow of symbols) called ``polytopic analysis of music'', and introduces an open-source toolbox dedicated to this scheme~\cite{PolytopeToolbox}.

The idea of linking music structure and compression schemes probably trace back to works such as Meyer' principles~\cite{meyer1956emotion}, Lerdahl \& Jackendoff Generative Theory of Tonal Music~\cite{lerdahl1983generative} and Narmour's Implication-Realization model~\cite{narmour1990analysis}. These works focus on music perception to capture some sense of music coherence in pieces, and are in that sense knowledge-based models.

On the other hand, probabilistic and Information-Theory-oriented models (hence, models which are less driven by prior knowledge) have been studied to capture structures of songs, such as the IDyOM (Information Dynamics Of
Music) which studies the Information Content of musical events~\cite{pearce2010melodic}, models based on the Kolmogorov complexity~\cite{meredith2012music}, or, more recently, Stochastic Neural Networks such as Restricted Boltzmann Machines~\cite{lattner2019modeling}.

An exhaustive list of compression-based structural segmentation models is beyond the scope of this article, and such a review can be found in~\cite{pearce2010melodic}. For its part, this article focuses on the S\&C model~\cite{bimbot2016system}, stemming from Narmour's theory~\cite{narmour1990analysis}.

This work is principally based on the previous work of C. Guichaoua~\cite{theseGuichaoua}, which was only presented in french until now. It is based on geometrical objects, called ``polytopes'', which support atomic musical elements on its vertices, and allows to study these musical elements in a non-chronological manner.

In that sense, polytopes can highlight repetition in music which don't occur sequentially, and aims at evaluating the information (or, more informally, the ``novelty'') brought by each musical element. Polytopes are well suited for the compression of musical information, as demonstrated in~\cite{theseLouboutin}, which shares this Information-Theory point of view to study repetitions and anticipations in music\footnote{As a matter of fact, the toolbox \textit{MusicOnPolytopes} also includes the work presented in~\cite{theseLouboutin}, and extends it for the task of structural segmentation by defining costs for irregular polytopes. Still, as both paradigms differ in numerous points, and for clarity, it is not presented here. An interesting reader should refer to~\cite{theseLouboutin}.}.

Polytopic analysis of music focuses on retrieving the frontiers between segments (\textit{i.e.}, structural boundaries), which are the time instances separating two consecutive parts. It does not study the labelling stage of segments, which consists of labelling in a same way coherent segments, and distinguishing dissimilar ones. A complete description of the structural segmentation task can be found in~\cite{nieto2020audio}.

This article presents a compression-base definition of the structural segmentation task, in Section~\ref{sec:compression}. Then, the main components of polytopic analysis of music are introduced in Section~\ref{sec:polytopes}, and the compression-oriented cost function associated with these objects is presented in Section~\ref{sec:guichaoua}. Finally, numerical experiments on the RWC Pop database are presented in Section~\ref{sec:experiences}.

\section{Structural Segmentation as a Compression Scheme}
\label{sec:compression}
\subsection{Definition of the Problem}
The polytopic analysis of music considers that the structure in music can be found by evaluating its internal repetitions, and by regrouping the similar passages in sections. Formally, this can be obtained in an optimization scheme, by defining the optimal structure as the structure of maximal compression, \textit{i.e.} the structure minimizing a complexity cost, left to be defined.

This work only considers the compression of music in a symbolic form, meaning that music is discretized both in time and features as a flow of symbols.

Practically, music is considered as discretized on musical beats\footnote{Other discretization could be considered, but musical beats has the advantage of being a musically-motivated discretization of time.}, and symbols represent the 24 major or minor perfect chords, summing up the musical content to the leading triad in the harmony. Let us denote $\{a_t, \, 1 \leq t \leq T\}$ this symbolic representation, $a_t$ being the symbol used to represent music at time $t$ (aligned with beats), and $T$ the length of the song.

The structural task now consists in finding a set of segments $Z = \{S_n, \, 1 \leq n \leq N\}$. Each $S_n$ is a sequence of consecutive elements $a_t$, such that $S_n = \{a_{t_n}, a_{t_n + 1}, ..., a_{t_{n+1} -1}\}$. The set $\{S_n\}$ partitions the $\{a_t\}$, in the sense that every $a_t$ belongs to one and only one $S_n$. Indexes $\{t_n\}$ are frontiers between segments.

Let's suppose for now the existence of a complexity cost function $\mathscr{C}$ applying on musical passages. The structural segmentation task is now defined as the search of the optimal sequence of segments regarding this cost, \textit{i.e.}, denoting as $\mathscr{C}(S_n)$ the cost of segment $S_n$:
\begin{equation}
    Z^* = \argmin_{Z} \sum_{n=1}^{N} \mathscr{C}(S_n)
\end{equation}
This is an optimization problem, which can be solved by a combinatorial analysis of the possible solutions. In particular, Sargent \textit{and al.}~\cite{sargent2016estimating} presented a dynamic programming algorithm which iteratively computes the optimal segmentation with respect to a cost.

The structural segmentation task is now reframed as the search of a complexity cost function $\mathscr{C}$ for segments, representing the main content of this article.

\subsection{Relation Between Musical Elements}
The development of this complexity cost function $\mathscr{C}$ is based on the study of relations between elements. Let's denote by $A$ the set of possible elements: $\forall 1 \leq i \leq T, a_i \in A$. In this work, $A$ represents all major and minor perfect chords, \textit{i.e.} $A = \llbracket0, 23\rrbracket$.

These elements are studied relatively with the others (and not individually). Precisely, let $G_r$ an abelian group of elements called relations. This group allows us to operate on $A$, which means that relations in $G_r$ act on the group $A$.

Hence:
\begin{itemize}
  \item $\forall f \in G_r, \, \forall a \in A, \, f.a \in A$,
  \item for all pair of elements $(a_i, a_j) \in A, \, \exists f \in G_r / f . a_i = a_j$.
  \end{itemize}
To simplify notations, we denote as $f_{i/j}$ the relation between $a_i$ and $a_j$. This group ensures that, for every musical passage $\{a_1, a_2, ..., a_n\}$, one can represent the relation between any two elements of this passage.

\subsubsection{Triad Circle}
A first set of relations, defined in~\cite{theseGuichaoua}, is the ``triad circle''. This circle is represented in Figure~\ref{fig:triad_circle}. Chords in this circle are ordered such that a clockwise rotation of one step represents the increase of the root by a third (\textit{i.e.} using respectively the third and the fifth of the first chord as the root and the third of the second chord).

The relation $f$ between 2 musical elements is defined as the number of steps between these two elements in the circle, or, differently said, the clockwise oriented angle between two elements. Hence, $\forall f \in G_f, f \in \llbracket-11, 12\rrbracket$.

\subsubsection{Tonnetz}
\begin{figure}[!ht]
  \begin{subfigure}[ht]{0.5\columnwidth}
 \centerline{\includegraphics[width=0.7\columnwidth]{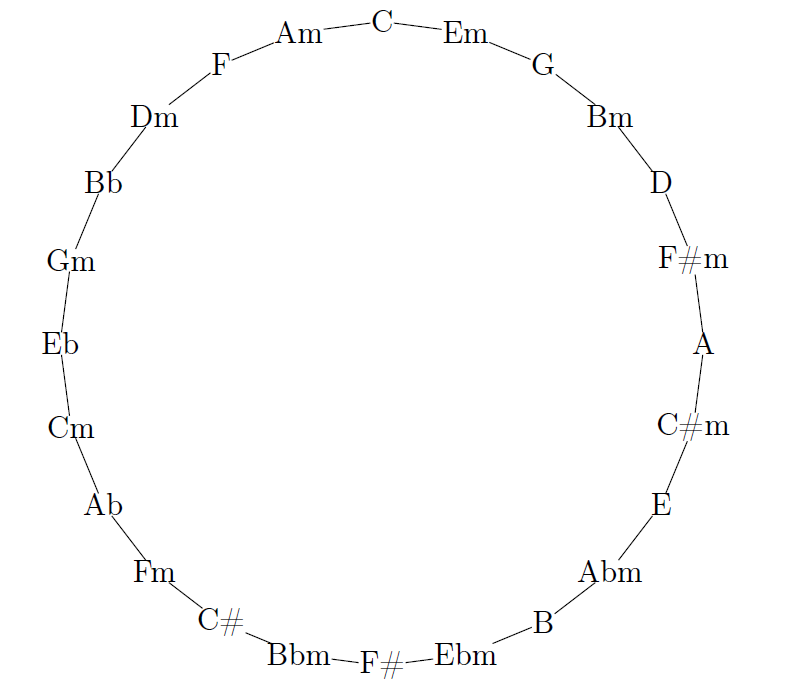}}
    \caption{Triad circle (from~\cite{theseGuichaoua}).}
    \label{fig:triad_circle}
\end{subfigure}
\quad
  \begin{subfigure}[ht]{0.5\columnwidth}
 \centerline{\includegraphics[width=\columnwidth]{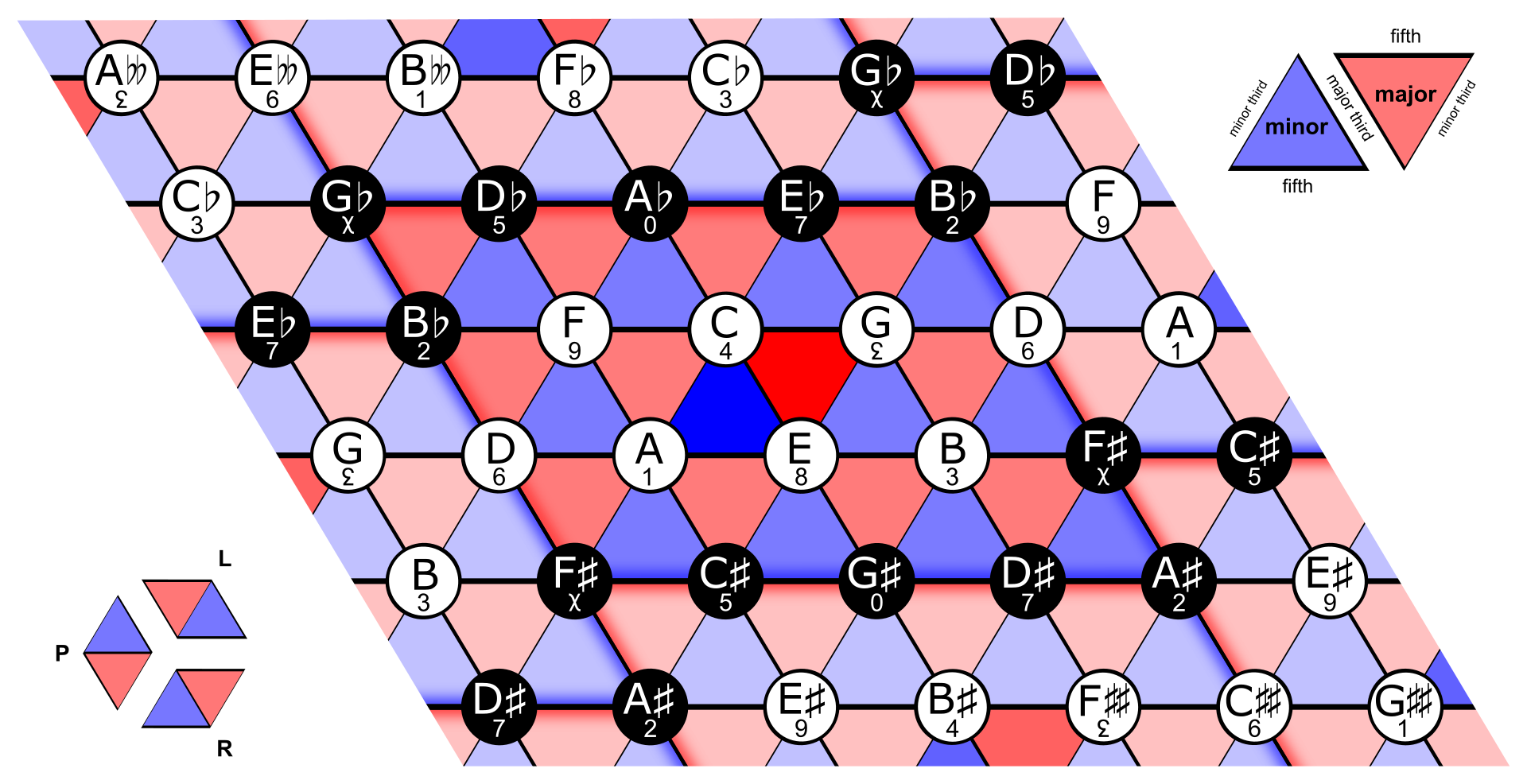}}
    \caption{Tonnetz \\(By Watchduck (a.k.a. Tilman Piesk) - Own work, CC0, https://commons.wikimedia.org/w/index.php?curid=33932849)}
    \label{fig:tonnetz}
\end{subfigure}
 \caption{Two models of relations: the triad circle and the Tonnetz.}
 \label{fig:relation_systems}
\end{figure}

A second set of relations is based on the Tonnetz, and more particularly the Neo-Riemannian Tonnetz subject to the western 12-chromatic scale~\cite{cohn1997neo}. The tonnetz is a lattice whose elements are ordered according to 3 harmonic relations between triads: P (Parallel, for triads sharing a common fifth, which represents here relations between a minor and a major triad sharing the same root, like A major and A minor), R (Relative, for triads sharing a common major third, such as A minor and C major) and L (for Leading-tone exchange, meaning that both triads share a common minor third, such as C major and E minor). It is represented in Figure \ref{fig:tonnetz}.

In this Tonnetz, two perfect chords can be compared as a sequence of composition of these three PLR relations. Even if the relation between two chords is not unique, one can define the ``canonic'' relation between them as the shortest relation in number of PLR relations.\footnote{Note that using the tonnetz with only the L and R relations redefines the previous triad circle.}

\section{Polytopic Analysis of Music}
\label{sec:polytopes}
\subsection{Polytopes}
A polytope is a geometrical pattern, composed of vertices and oriented edges (arrows). Polytopes are defined to scale up the previously defined relations to musical passages. Vertices and arrows of a polytope respectively represent musical elements $a_t$ and their relations $f$.

\begin{definition}[Regular polytope] Primary polytopes are n-dimensional hypercubes. They are of the form of a square, a cube, a tesseract, etc.
\end{definition}

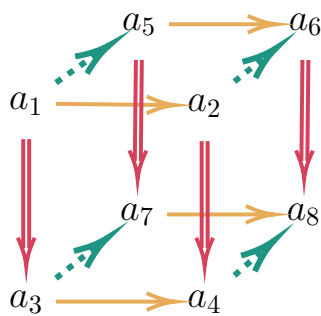
\begin{figure}[ht]
    \centering
\begin{tikzpicture}[x=0.75pt,y=0.75pt,yscale=-1,xscale=1]

\draw (45,126.33) node  [font=\Large]  {$a_{1}$};
\draw (102,86.33) node  [font=\Large]  {$a_{5}$};
\draw (45,226.33) node  [font=\Large]  {$a_{3}$};
\draw (186,86.33) node  [font=\Large]  {$a_{6}$};
\draw (135,127.33) node  [font=\Large]  {$a_{2}$};
\draw (101,182.33) node  [font=\Large]  {$a_{7}$};
\draw (185,182.33) node  [font=\Large]  {$a_{8}$};
\draw (135,226.33) node  [font=\Large]  {$a_{4}$};
\draw [color={rgb, 255:red, 230; green, 172; blue, 90 }  ,draw opacity=1 ][line width=1.5]    (117,86.33) -- (168,86.33) ;
\draw [shift={(171,86.33)}, rotate = 180] [color={rgb, 255:red, 230; green, 172; blue, 90 }  ,draw opacity=1 ][line width=1.5]    (14.21,-4.28) .. controls (9.04,-1.82) and (4.3,-0.39) .. (0,0) .. controls (4.3,0.39) and (9.04,1.82) .. (14.21,4.28)   ;
\draw [color={rgb, 255:red, 32; green, 148; blue, 131 }  ,draw opacity=1 ][line width=2.25]  [dash pattern={on 2.53pt off 3.02pt}]  (60,115.81) -- (83.73,99.16) ;
\draw [shift={(87,96.86)}, rotate = 504.94] [color={rgb, 255:red, 32; green, 148; blue, 131 }  ,draw opacity=1 ][line width=2.25]    (17.49,-5.26) .. controls (11.12,-2.23) and (5.29,-0.48) .. (0,0) .. controls (5.29,0.48) and (11.12,2.23) .. (17.49,5.26)   ;
\draw [color={rgb, 255:red, 230; green, 172; blue, 90 }  ,draw opacity=1 ][line width=1.5]    (60,126.5) -- (117,127.13) ;
\draw [shift={(120,127.17)}, rotate = 180.64] [color={rgb, 255:red, 230; green, 172; blue, 90 }  ,draw opacity=1 ][line width=1.5]    (14.21,-4.28) .. controls (9.04,-1.82) and (4.3,-0.39) .. (0,0) .. controls (4.3,0.39) and (9.04,1.82) .. (14.21,4.28)   ;
\draw [color={rgb, 255:red, 32; green, 148; blue, 131 }  ,draw opacity=1 ][line width=2.25]  [dash pattern={on 2.53pt off 3.02pt}]  (150,115.27) -- (167.88,100.9) ;
\draw [shift={(171,98.39)}, rotate = 501.2] [color={rgb, 255:red, 32; green, 148; blue, 131 }  ,draw opacity=1 ][line width=2.25]    (17.49,-5.26) .. controls (11.12,-2.23) and (5.29,-0.48) .. (0,0) .. controls (5.29,0.48) and (11.12,2.23) .. (17.49,5.26)   ;
\draw [color={rgb, 255:red, 213; green, 65; blue, 90 }  ,draw opacity=1 ][line width=1.5]    (46.5,144.33) -- (46.5,199.33)(43.5,144.33) -- (43.5,199.33) ;
\draw [shift={(45,208.33)}, rotate = 270] [color={rgb, 255:red, 213; green, 65; blue, 90 }  ,draw opacity=1 ][line width=1.5]    (14.21,-4.28) .. controls (9.04,-1.82) and (4.3,-0.39) .. (0,0) .. controls (4.3,0.39) and (9.04,1.82) .. (14.21,4.28)   ;
\draw [color={rgb, 255:red, 230; green, 172; blue, 90 }  ,draw opacity=1 ][line width=1.5]    (60,226.33) -- (117,226.33) ;
\draw [shift={(120,226.33)}, rotate = 180] [color={rgb, 255:red, 230; green, 172; blue, 90 }  ,draw opacity=1 ][line width=1.5]    (14.21,-4.28) .. controls (9.04,-1.82) and (4.3,-0.39) .. (0,0) .. controls (4.3,0.39) and (9.04,1.82) .. (14.21,4.28)   ;
\draw [color={rgb, 255:red, 32; green, 148; blue, 131 }  ,draw opacity=1 ][line width=2.25]  [dash pattern={on 2.53pt off 3.02pt}]  (150,213.13) -- (167,198.18) ;
\draw [shift={(170,195.53)}, rotate = 498.65] [color={rgb, 255:red, 32; green, 148; blue, 131 }  ,draw opacity=1 ][line width=2.25]    (17.49,-5.26) .. controls (11.12,-2.23) and (5.29,-0.48) .. (0,0) .. controls (5.29,0.48) and (11.12,2.23) .. (17.49,5.26)   ;
\draw [color={rgb, 255:red, 32; green, 148; blue, 131 }  ,draw opacity=1 ][line width=2.25]  [dash pattern={on 2.53pt off 3.02pt}]  (60,214.55) -- (82.85,196.59) ;
\draw [shift={(86,194.12)}, rotate = 501.84] [color={rgb, 255:red, 32; green, 148; blue, 131 }  ,draw opacity=1 ][line width=2.25]    (17.49,-5.26) .. controls (11.12,-2.23) and (5.29,-0.48) .. (0,0) .. controls (5.29,0.48) and (11.12,2.23) .. (17.49,5.26)   ;
\draw [color={rgb, 255:red, 230; green, 172; blue, 90 }  ,draw opacity=1 ][line width=1.5]    (116,182.33) -- (167,182.33) ;
\draw [shift={(170,182.33)}, rotate = 180] [color={rgb, 255:red, 230; green, 172; blue, 90 }  ,draw opacity=1 ][line width=1.5]    (14.21,-4.28) .. controls (9.04,-1.82) and (4.3,-0.39) .. (0,0) .. controls (4.3,0.39) and (9.04,1.82) .. (14.21,4.28)   ;
\draw [color={rgb, 255:red, 213; green, 65; blue, 90 }  ,draw opacity=1 ][line width=1.5]    (103.31,104.35) -- (102.78,155.35)(100.31,104.32) -- (99.78,155.32) ;
\draw [shift={(101.19,164.33)}, rotate = 270.6] [color={rgb, 255:red, 213; green, 65; blue, 90 }  ,draw opacity=1 ][line width=1.5]    (14.21,-4.28) .. controls (9.04,-1.82) and (4.3,-0.39) .. (0,0) .. controls (4.3,0.39) and (9.04,1.82) .. (14.21,4.28)   ;
\draw [color={rgb, 255:red, 213; green, 65; blue, 90 }  ,draw opacity=1 ][line width=1.5]    (187.31,104.35) -- (186.78,155.35)(184.31,104.32) -- (183.78,155.32) ;
\draw [shift={(185.19,164.33)}, rotate = 270.6] [color={rgb, 255:red, 213; green, 65; blue, 90 }  ,draw opacity=1 ][line width=1.5]    (14.21,-4.28) .. controls (9.04,-1.82) and (4.3,-0.39) .. (0,0) .. controls (4.3,0.39) and (9.04,1.82) .. (14.21,4.28)   ;
\draw [color={rgb, 255:red, 213; green, 65; blue, 90 }  ,draw opacity=1 ][line width=1.5]    (136.5,145.33) -- (136.5,199.33)(133.5,145.33) -- (133.5,199.33) ;
\draw [shift={(135,208.33)}, rotate = 270] [color={rgb, 255:red, 213; green, 65; blue, 90 }  ,draw opacity=1 ][line width=1.5]    (14.21,-4.28) .. controls (9.04,-1.82) and (4.3,-0.39) .. (0,0) .. controls (4.3,0.39) and (9.04,1.82) .. (14.21,4.28)   ;

\end{tikzpicture}

    \caption{3-polytope (cube)}
    \label{fig:3_dim_polytope}
\end{figure}

A n-dimensional regular polytope is defined by its \textbf{dimension}: a regular n-dimensional polytope contains $2^n$ elements. Hence, a 2-dimensional regular polytope represents a square and contains 4 elements; a 3-dimensional regular polytope contains 8 elements and represents a cube; etc. A 3-dimensional regular polytope (called 3-polytope for simplification) is represented in Figure~\ref{fig:3_dim_polytope}.

Hence, regular polytopes can model musical passages of $2^n$ elements, but are not suited for passages of different sizes. In order to consider a large number of passages size, we extend these regular polytopes to ``irregular polytopes''.

\begin{definition}[Irregular polytope] A n-dimensional irregular polytope correspond to a n-dimensional regular polytope on which has been deleted and/or added some vertices and edges. These alteration (either addition or deletion) follow themselves the shape of a regular polytope of dimension $d < n-1$, \textit{i.e.} deleted and/or added vertices form themselves a d-dimensional regular polytope.

  An irregular polytope is constructed from at most one d-polytope modeling the addition and at most d'-polytope representing the deletion.
\end{definition}

For example, starting from a 3-dimensional regular polytope (a cube), and deleting its last vertex (0-polytope) results in a 3-dimensional irregular polytope with 7 elements instead of 8. To detail the construction specifications of these irregular polytopes, we further introduce the notions of \textbf{antecedent} and \textbf{successor}.

\begin{definition}[Antecedent] Let an element be the extremity of (at least) one arrow. We define the antecedent(s) of this element as the origin(s) of this (or these) arrow(s). An element can have several antecedents if it is the extremity of several arrows. In Figure~\ref{fig:3_dim_polytope}, $a_2$ and $a_3$ are two antecedents of $a_4$.
\end{definition}

\begin{definition}[Successor] Let an element be the origin of (at least) one arrow. We define the successors of this element as all elements which are at the extremity of this (or these) arrow(s). Elements do not necessarily have successors. In Figure \ref{fig:3_dim_polytope}, $a_4$ and $a_7$ are successors of $a_3$.
\end{definition}

As an edge represents a relation between two elements, every edge must connect existing elements. Hence, deleting a vertex implies the deletion of all arrows starting from its antecedents.

In addition, deleting an element at the origin of an edge implies the deletion of this edge, which can result in elements without arrows connecting them to the polytope. Hence, deleting an element must imply the deletion of its successors.

To ensure this latter condition, every alteration polytope must include the last element of the polytope, and, when both addition and deletion operate on a same vertex, this vertex is deleted without addition (\textit{i.e.} deletion is preferred over addition).

Similarly, every added element must be connected to another element of the polytope by an edge. In that sense, when adding an element, an edge is created with the vertex supporting this addition. This new element is considered as ``attached'' to this vertex. As the additional edges form themselves a polytope, added elements are connected by new edges.

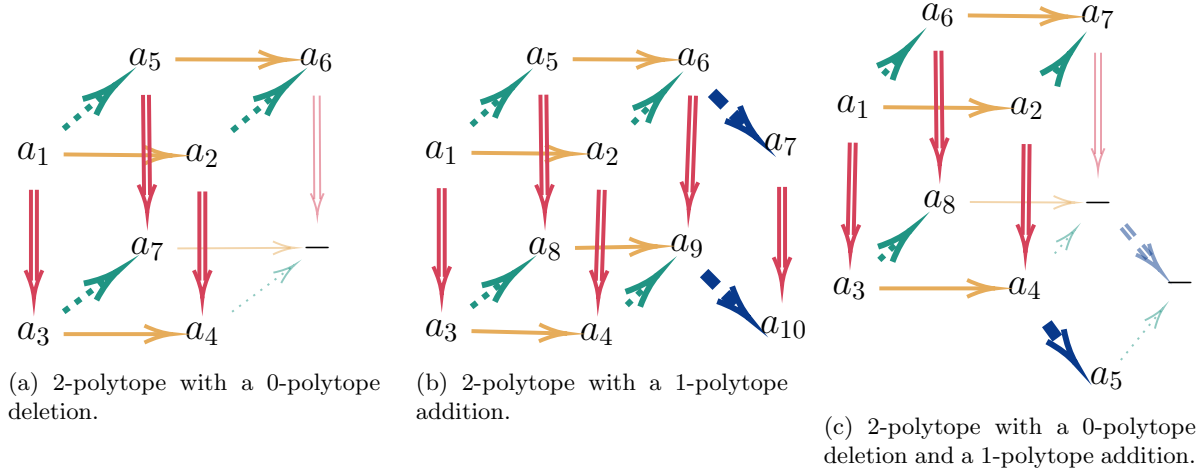
\begin{figure}[!ht]
  \begin{subfigure}[ht]{0.3\columnwidth}
\begin{tikzpicture}[x=0.75pt,y=0.75pt,yscale=-1,xscale=1]

\draw (19.74,112.55) node  [font=\Large]  {$a_{1}$};
\draw (76.07,64.65) node  [font=\Large]  {$a_{5}$};
\draw (19.74,203.4) node  [font=\Large]  {$a_{3}$};
\draw (162.28,65) node  [font=\Large]  {$a_{6}$};
\draw (104.68,113.41) node  [font=\Large]  {$a_{2}$};
\draw (77.07,160.23) node  [font=\Large]  {$a_{7}$};
\draw (162.74,159.79) node  [font=\Large]  {$-$};
\draw (104.68,203.4) node  [font=\Large]  {$a_{4}$};
\draw [color={rgb, 255:red, 230; green, 172; blue, 90 }  ,draw opacity=1 ][line width=1.5]    (91.07,64.71) -- (144.28,64.93) ;
\draw [shift={(147.28,64.94)}, rotate = 180.23] [color={rgb, 255:red, 230; green, 172; blue, 90 }  ,draw opacity=1 ][line width=1.5]    (14.21,-4.28) .. controls (9.04,-1.82) and (4.3,-0.39) .. (0,0) .. controls (4.3,0.39) and (9.04,1.82) .. (14.21,4.28)   ;
\draw [color={rgb, 255:red, 32; green, 148; blue, 131 }  ,draw opacity=1 ][line width=2.25]  [dash pattern={on 2.53pt off 3.02pt}]  (34.74,99.79) -- (58.02,80) ;
\draw [shift={(61.07,77.41)}, rotate = 499.63] [color={rgb, 255:red, 32; green, 148; blue, 131 }  ,draw opacity=1 ][line width=2.25]    (17.49,-5.26) .. controls (11.12,-2.23) and (5.29,-0.48) .. (0,0) .. controls (5.29,0.48) and (11.12,2.23) .. (17.49,5.26)   ;
\draw [color={rgb, 255:red, 230; green, 172; blue, 90 }  ,draw opacity=1 ][line width=1.5]    (34.74,112.7) -- (86.68,113.23) ;
\draw [shift={(89.68,113.26)}, rotate = 180.59] [color={rgb, 255:red, 230; green, 172; blue, 90 }  ,draw opacity=1 ][line width=1.5]    (14.21,-4.28) .. controls (9.04,-1.82) and (4.3,-0.39) .. (0,0) .. controls (4.3,0.39) and (9.04,1.82) .. (14.21,4.28)   ;
\draw [color={rgb, 255:red, 32; green, 148; blue, 131 }  ,draw opacity=1 ][line width=2.25]  [dash pattern={on 2.53pt off 3.02pt}]  (119.68,100.81) -- (144.22,80.18) ;
\draw [shift={(147.28,77.61)}, rotate = 499.96] [color={rgb, 255:red, 32; green, 148; blue, 131 }  ,draw opacity=1 ][line width=2.25]    (17.49,-5.26) .. controls (11.12,-2.23) and (5.29,-0.48) .. (0,0) .. controls (5.29,0.48) and (11.12,2.23) .. (17.49,5.26)   ;
\draw [color={rgb, 255:red, 213; green, 65; blue, 90 }  ,draw opacity=1 ][line width=1.5]    (21.24,130.55) -- (21.24,176.4)(18.24,130.55) -- (18.24,176.4) ;
\draw [shift={(19.74,185.4)}, rotate = 270] [color={rgb, 255:red, 213; green, 65; blue, 90 }  ,draw opacity=1 ][line width=1.5]    (14.21,-4.28) .. controls (9.04,-1.82) and (4.3,-0.39) .. (0,0) .. controls (4.3,0.39) and (9.04,1.82) .. (14.21,4.28)   ;
\draw [color={rgb, 255:red, 230; green, 172; blue, 90 }  ,draw opacity=1 ][line width=1.5]    (34.74,203.4) -- (86.68,203.4) ;
\draw [shift={(89.68,203.4)}, rotate = 180] [color={rgb, 255:red, 230; green, 172; blue, 90 }  ,draw opacity=1 ][line width=1.5]    (14.21,-4.28) .. controls (9.04,-1.82) and (4.3,-0.39) .. (0,0) .. controls (4.3,0.39) and (9.04,1.82) .. (14.21,4.28)   ;
\draw [color={rgb, 255:red, 32; green, 148; blue, 131 }  ,draw opacity=0.5 ][line width=0.75]  [dash pattern={on 0.84pt off 2.51pt}]  (119.68,192.13) -- (149.64,169.63) ;
\draw [shift={(151.24,168.43)}, rotate = 503.09] [color={rgb, 255:red, 32; green, 148; blue, 131 }  ,draw opacity=0.5 ][line width=0.75]    (10.93,-3.29) .. controls (6.95,-1.4) and (3.31,-0.3) .. (0,0) .. controls (3.31,0.3) and (6.95,1.4) .. (10.93,3.29)   ;
\draw [color={rgb, 255:red, 32; green, 148; blue, 131 }  ,draw opacity=1 ][line width=2.25]  [dash pattern={on 2.53pt off 3.02pt}]  (34.74,192.1) -- (58.87,173.93) ;
\draw [shift={(62.07,171.53)}, rotate = 503.02] [color={rgb, 255:red, 32; green, 148; blue, 131 }  ,draw opacity=1 ][line width=2.25]    (17.49,-5.26) .. controls (11.12,-2.23) and (5.29,-0.48) .. (0,0) .. controls (5.29,0.48) and (11.12,2.23) .. (17.49,5.26)   ;
\draw [color={rgb, 255:red, 230; green, 172; blue, 90 }  ,draw opacity=0.5 ][line width=0.75]    (92.07,160.15) -- (149.24,159.86) ;
\draw [shift={(151.24,159.85)}, rotate = 539.71] [color={rgb, 255:red, 230; green, 172; blue, 90 }  ,draw opacity=0.5 ][line width=0.75]    (10.93,-3.29) .. controls (6.95,-1.4) and (3.31,-0.3) .. (0,0) .. controls (3.31,0.3) and (6.95,1.4) .. (10.93,3.29)   ;
\draw [color={rgb, 255:red, 213; green, 65; blue, 90 }  ,draw opacity=1 ][line width=1.5]    (77.76,82.64) -- (78.29,133.22)(74.76,82.67) -- (75.29,133.25) ;
\draw [shift={(76.88,142.23)}, rotate = 269.4] [color={rgb, 255:red, 213; green, 65; blue, 90 }  ,draw opacity=1 ][line width=1.5]    (14.21,-4.28) .. controls (9.04,-1.82) and (4.3,-0.39) .. (0,0) .. controls (4.3,0.39) and (9.04,1.82) .. (14.21,4.28)   ;
\draw [color={rgb, 255:red, 213; green, 65; blue, 90 }  ,draw opacity=0.5 ][line width=0.75]    (163.87,83) -- (164.12,135.28)(160.87,83.01) -- (161.12,135.3) ;
\draw [shift={(162.66,143.29)}, rotate = 269.72] [color={rgb, 255:red, 213; green, 65; blue, 90 }  ,draw opacity=0.5 ][line width=0.75]    (10.93,-3.29) .. controls (6.95,-1.4) and (3.31,-0.3) .. (0,0) .. controls (3.31,0.3) and (6.95,1.4) .. (10.93,3.29)   ;
\draw [color={rgb, 255:red, 213; green, 65; blue, 90 }  ,draw opacity=1 ][line width=1.5]    (106.18,131.41) -- (106.18,176.4)(103.18,131.41) -- (103.18,176.4) ;
\draw [shift={(104.68,185.4)}, rotate = 270] [color={rgb, 255:red, 213; green, 65; blue, 90 }  ,draw opacity=1 ][line width=1.5]    (14.21,-4.28) .. controls (9.04,-1.82) and (4.3,-0.39) .. (0,0) .. controls (4.3,0.39) and (9.04,1.82) .. (14.21,4.28)   ;

\end{tikzpicture}

\caption{2-polytope with a 0-polytope deletion.}
    \label{fig:deletion_subexample}
\end{subfigure}
\quad
  \begin{subfigure}[ht]{0.3\columnwidth}
\begin{tikzpicture}[x=0.75pt,y=0.75pt,yscale=-1,xscale=1]

\draw [color={rgb, 255:red, 8; green, 57; blue, 139 }  ,draw opacity=1 ][line width=2.25]  [dash pattern={on 6.75pt off 4.5pt}]  (374.47,82.33) -- (390.14,98.09)(372.34,84.45) -- (388.02,100.2) ;
\draw [shift={(396.13,106.23)}, rotate = 225.14] [color={rgb, 255:red, 8; green, 57; blue, 139 }  ,draw opacity=1 ][line width=2.25]    (17.49,-5.26) .. controls (11.12,-2.23) and (5.29,-0.48) .. (0,0) .. controls (5.29,0.48) and (11.12,2.23) .. (17.49,5.26)   ;
\draw [color={rgb, 255:red, 213; green, 65; blue, 90 }  ,draw opacity=1 ][line width=1.5]    (410.21,130.87) -- (410.21,171.61)(407.21,130.87) -- (407.21,171.61) ;
\draw [shift={(408.71,180.61)}, rotate = 270] [color={rgb, 255:red, 213; green, 65; blue, 90 }  ,draw opacity=1 ][line width=1.5]    (14.21,-4.28) .. controls (9.04,-1.82) and (4.3,-0.39) .. (0,0) .. controls (4.3,0.39) and (9.04,1.82) .. (14.21,4.28)   ;
\draw [color={rgb, 255:red, 8; green, 57; blue, 139 }  ,draw opacity=1 ][line width=2.25]  [dash pattern={on 6.75pt off 4.5pt}]  (370.47,173.33) -- (386.14,189.09)(368.34,175.45) -- (384.02,191.2) ;
\draw [shift={(392.13,197.23)}, rotate = 225.14] [color={rgb, 255:red, 8; green, 57; blue, 139 }  ,draw opacity=1 ][line width=2.25]    (17.49,-5.26) .. controls (11.12,-2.23) and (5.29,-0.48) .. (0,0) .. controls (5.29,0.48) and (11.12,2.23) .. (17.49,5.26)   ;

\draw (237.07,113.51) node  [font=\Large]  {$a_{1}$};
\draw (288.12,66.34) node  [font=\Large]  {$a_{5}$};
\draw (237.07,203.27) node  [font=\Large]  {$a_{3}$};
\draw (364.48,66.8) node  [font=\Large]  {$a_{6}$};
\draw (318.4,114.34) node  [font=\Large]  {$a_{2}$};
\draw (289.28,161.97) node  [font=\Large]  {$a_{8}$};
\draw (360.95,160.63) node  [font=\Large]  {$a_{9}$};
\draw (315.4,205.27) node  [font=\Large]  {$a_{4}$};
\draw (409.53,201.74) node  [font=\Large]  {$a_{10}$};
\draw (408.2,109.34) node  [font=\Large]  {$a_{7}$};
\draw [color={rgb, 255:red, 230; green, 172; blue, 90 }  ,draw opacity=1 ][line width=1.5]    (303.12,66.43) -- (346.48,66.69) ;
\draw [shift={(349.48,66.71)}, rotate = 180.35] [color={rgb, 255:red, 230; green, 172; blue, 90 }  ,draw opacity=1 ][line width=1.5]    (14.21,-4.28) .. controls (9.04,-1.82) and (4.3,-0.39) .. (0,0) .. controls (4.3,0.39) and (9.04,1.82) .. (14.21,4.28)   ;
\draw [color={rgb, 255:red, 32; green, 148; blue, 131 }  ,draw opacity=1 ][line width=2.25]  [dash pattern={on 2.53pt off 3.02pt}]  (252.07,99.65) -- (270.19,82.91) ;
\draw [shift={(273.12,80.2)}, rotate = 497.26] [color={rgb, 255:red, 32; green, 148; blue, 131 }  ,draw opacity=1 ][line width=2.25]    (17.49,-5.26) .. controls (11.12,-2.23) and (5.29,-0.48) .. (0,0) .. controls (5.29,0.48) and (11.12,2.23) .. (17.49,5.26)   ;
\draw [color={rgb, 255:red, 230; green, 172; blue, 90 }  ,draw opacity=1 ][line width=1.5]    (252.07,113.67) -- (300.4,114.16) ;
\draw [shift={(303.4,114.19)}, rotate = 180.58] [color={rgb, 255:red, 230; green, 172; blue, 90 }  ,draw opacity=1 ][line width=1.5]    (14.21,-4.28) .. controls (9.04,-1.82) and (4.3,-0.39) .. (0,0) .. controls (4.3,0.39) and (9.04,1.82) .. (14.21,4.28)   ;
\draw [color={rgb, 255:red, 32; green, 148; blue, 131 }  ,draw opacity=1 ][line width=2.25]  [dash pattern={on 2.53pt off 3.02pt}]  (333.4,98.87) -- (346.7,85.15) ;
\draw [shift={(349.48,82.28)}, rotate = 494.11] [color={rgb, 255:red, 32; green, 148; blue, 131 }  ,draw opacity=1 ][line width=2.25]    (17.49,-5.26) .. controls (11.12,-2.23) and (5.29,-0.48) .. (0,0) .. controls (5.29,0.48) and (11.12,2.23) .. (17.49,5.26)   ;
\draw [color={rgb, 255:red, 213; green, 65; blue, 90 }  ,draw opacity=1 ][line width=1.5]    (238.57,131.51) -- (238.57,176.27)(235.57,131.51) -- (235.57,176.27) ;
\draw [shift={(237.07,185.27)}, rotate = 270] [color={rgb, 255:red, 213; green, 65; blue, 90 }  ,draw opacity=1 ][line width=1.5]    (14.21,-4.28) .. controls (9.04,-1.82) and (4.3,-0.39) .. (0,0) .. controls (4.3,0.39) and (9.04,1.82) .. (14.21,4.28)   ;
\draw [color={rgb, 255:red, 230; green, 172; blue, 90 }  ,draw opacity=1 ][line width=1.5]    (252.07,203.66) -- (297.4,204.81) ;
\draw [shift={(300.4,204.89)}, rotate = 181.46] [color={rgb, 255:red, 230; green, 172; blue, 90 }  ,draw opacity=1 ][line width=1.5]    (14.21,-4.28) .. controls (9.04,-1.82) and (4.3,-0.39) .. (0,0) .. controls (4.3,0.39) and (9.04,1.82) .. (14.21,4.28)   ;
\draw [color={rgb, 255:red, 32; green, 148; blue, 131 }  ,draw opacity=1 ][line width=2.25]  [dash pattern={on 2.53pt off 3.02pt}]  (330.4,190.57) -- (343.1,178.13) ;
\draw [shift={(345.95,175.33)}, rotate = 495.58] [color={rgb, 255:red, 32; green, 148; blue, 131 }  ,draw opacity=1 ][line width=2.25]    (17.49,-5.26) .. controls (11.12,-2.23) and (5.29,-0.48) .. (0,0) .. controls (5.29,0.48) and (11.12,2.23) .. (17.49,5.26)   ;
\draw [color={rgb, 255:red, 32; green, 148; blue, 131 }  ,draw opacity=1 ][line width=2.25]  [dash pattern={on 2.53pt off 3.02pt}]  (252.07,191.41) -- (271.14,176.32) ;
\draw [shift={(274.28,173.84)}, rotate = 501.65] [color={rgb, 255:red, 32; green, 148; blue, 131 }  ,draw opacity=1 ][line width=2.25]    (17.49,-5.26) .. controls (11.12,-2.23) and (5.29,-0.48) .. (0,0) .. controls (5.29,0.48) and (11.12,2.23) .. (17.49,5.26)   ;
\draw [color={rgb, 255:red, 230; green, 172; blue, 90 }  ,draw opacity=1 ][line width=1.5]    (304.28,161.69) -- (342.96,160.97) ;
\draw [shift={(345.95,160.91)}, rotate = 538.9300000000001] [color={rgb, 255:red, 230; green, 172; blue, 90 }  ,draw opacity=1 ][line width=1.5]    (14.21,-4.28) .. controls (9.04,-1.82) and (4.3,-0.39) .. (0,0) .. controls (4.3,0.39) and (9.04,1.82) .. (14.21,4.28)   ;
\draw [color={rgb, 255:red, 213; green, 65; blue, 90 }  ,draw opacity=1 ][line width=1.5]    (289.84,84.32) -- (290.45,134.96)(286.84,84.35) -- (287.45,134.99) ;
\draw [shift={(289.06,143.97)}, rotate = 269.31] [color={rgb, 255:red, 213; green, 65; blue, 90 }  ,draw opacity=1 ][line width=1.5]    (14.21,-4.28) .. controls (9.04,-1.82) and (4.3,-0.39) .. (0,0) .. controls (4.3,0.39) and (9.04,1.82) .. (14.21,4.28)   ;
\draw [color={rgb, 255:red, 213; green, 65; blue, 90 }  ,draw opacity=1 ][line width=1.5]    (365.3,84.86) -- (363.47,133.69)(362.31,84.75) -- (360.47,133.58) ;
\draw [shift={(361.63,142.63)}, rotate = 272.15] [color={rgb, 255:red, 213; green, 65; blue, 90 }  ,draw opacity=1 ][line width=1.5]    (14.21,-4.28) .. controls (9.04,-1.82) and (4.3,-0.39) .. (0,0) .. controls (4.3,0.39) and (9.04,1.82) .. (14.21,4.28)   ;
\draw [color={rgb, 255:red, 213; green, 65; blue, 90 }  ,draw opacity=1 ][line width=1.5]    (319.3,132.39) -- (317.79,178.33)(316.31,132.29) -- (314.79,178.23) ;
\draw [shift={(315.99,187.27)}, rotate = 271.89] [color={rgb, 255:red, 213; green, 65; blue, 90 }  ,draw opacity=1 ][line width=1.5]    (14.21,-4.28) .. controls (9.04,-1.82) and (4.3,-0.39) .. (0,0) .. controls (4.3,0.39) and (9.04,1.82) .. (14.21,4.28)   ;

\end{tikzpicture}

\caption{2-polytope with a 1-polytope addition.}
    \label{fig:addition_subexample}
\end{subfigure}
\quad
  \begin{subfigure}[ht]{0.3\columnwidth}
\begin{tikzpicture}[x=0.75pt,y=0.75pt,yscale=-1,xscale=1]

\draw [color={rgb, 255:red, 8; green, 57; blue, 139 }  ,draw opacity=1 ][line width=2.25]  [dash pattern={on 6.75pt off 4.5pt}]  (582,221.03) -- (590.72,233.02)(579.57,222.8) -- (588.29,234.78) ;
\draw [shift={(595.39,241.99)}, rotate = 233.96] [color={rgb, 255:red, 8; green, 57; blue, 139 }  ,draw opacity=1 ][line width=2.25]    (17.49,-5.26) .. controls (11.12,-2.23) and (5.29,-0.48) .. (0,0) .. controls (5.29,0.48) and (11.12,2.23) .. (17.49,5.26)   ;
\draw [color={rgb, 255:red, 8; green, 57; blue, 139 }  ,draw opacity=0.5 ][line width=1.5]  [dash pattern={on 5.63pt off 4.5pt}]  (620.57,171.74) -- (633.54,187.35)(618.26,173.65) -- (631.23,189.27) ;
\draw [shift={(638.13,195.23)}, rotate = 230.29] [color={rgb, 255:red, 8; green, 57; blue, 139 }  ,draw opacity=0.5 ][line width=1.5]    (14.21,-4.28) .. controls (9.04,-1.82) and (4.3,-0.39) .. (0,0) .. controls (4.3,0.39) and (9.04,1.82) .. (14.21,4.28)   ;
\draw [color={rgb, 255:red, 32; green, 148; blue, 131 }  ,draw opacity=0.5 ][line width=0.75]  [dash pattern={on 0.84pt off 2.51pt}]  (618.13,240.23) -- (638.4,215.38) ;
\draw [shift={(639.67,213.83)}, rotate = 489.2] [color={rgb, 255:red, 32; green, 148; blue, 131 }  ,draw opacity=0.5 ][line width=0.75]    (10.93,-3.29) .. controls (6.95,-1.4) and (3.31,-0.3) .. (0,0) .. controls (3.31,0.3) and (6.95,1.4) .. (10.93,3.29)   ;

\draw (483.31,112.52) node  [font=\Large]  {$a_{1}$};
\draw (525.96,66.13) node  [font=\Large]  {$a_{6}$};
\draw (481.31,204.09) node  [font=\Large]  {$a_{3}$};
\draw (606.62,67.13) node  [font=\Large]  {$a_{7}$};
\draw (570.8,113.23) node  [font=\Large]  {$a_{2}$};
\draw (526.96,160.4) node  [font=\Large]  {$a_{8}$};
\draw (606.62,160.83) node  [font=\Large]  {$-$};
\draw (569.8,204.09) node  [font=\Large]  {$a_{4}$};
\draw (647.9,200.93) node  [font=\Large]  {$-$};
\draw (611.14,249.45) node  [font=\Large]  {$a_{5}$};
\draw [color={rgb, 255:red, 230; green, 172; blue, 90 }  ,draw opacity=1 ][line width=1.5]    (540.96,66.31) -- (588.62,66.9) ;
\draw [shift={(591.62,66.94)}, rotate = 180.71] [color={rgb, 255:red, 230; green, 172; blue, 90 }  ,draw opacity=1 ][line width=1.5]    (14.21,-4.28) .. controls (9.04,-1.82) and (4.3,-0.39) .. (0,0) .. controls (4.3,0.39) and (9.04,1.82) .. (14.21,4.28)   ;
\draw [color={rgb, 255:red, 32; green, 148; blue, 131 }  ,draw opacity=1 ][line width=2.25]  [dash pattern={on 2.53pt off 3.02pt}]  (498.31,96.2) -- (508.25,85.39) ;
\draw [shift={(510.96,82.44)}, rotate = 492.59] [color={rgb, 255:red, 32; green, 148; blue, 131 }  ,draw opacity=1 ][line width=2.25]    (17.49,-5.26) .. controls (11.12,-2.23) and (5.29,-0.48) .. (0,0) .. controls (5.29,0.48) and (11.12,2.23) .. (17.49,5.26)   ;
\draw [color={rgb, 255:red, 230; green, 172; blue, 90 }  ,draw opacity=1 ][line width=1.5]    (498.31,112.64) -- (552.8,113.09) ;
\draw [shift={(555.8,113.11)}, rotate = 180.47] [color={rgb, 255:red, 230; green, 172; blue, 90 }  ,draw opacity=1 ][line width=1.5]    (14.21,-4.28) .. controls (9.04,-1.82) and (4.3,-0.39) .. (0,0) .. controls (4.3,0.39) and (9.04,1.82) .. (14.21,4.28)   ;
\draw [color={rgb, 255:red, 32; green, 148; blue, 131 }  ,draw opacity=1 ][line width=2.25]  [dash pattern={on 2.53pt off 3.02pt}]  (584.78,95.23) -- (590.18,88.29) ;
\draw [shift={(592.63,85.13)}, rotate = 487.84] [color={rgb, 255:red, 32; green, 148; blue, 131 }  ,draw opacity=1 ][line width=2.25]    (17.49,-5.26) .. controls (11.12,-2.23) and (5.29,-0.48) .. (0,0) .. controls (5.29,0.48) and (11.12,2.23) .. (17.49,5.26)   ;
\draw [color={rgb, 255:red, 213; green, 65; blue, 90 }  ,draw opacity=1 ][line width=1.5]    (484.42,130.55) -- (483.4,177.12)(481.42,130.49) -- (480.4,177.06) ;
\draw [shift={(481.7,186.09)}, rotate = 271.25] [color={rgb, 255:red, 213; green, 65; blue, 90 }  ,draw opacity=1 ][line width=1.5]    (14.21,-4.28) .. controls (9.04,-1.82) and (4.3,-0.39) .. (0,0) .. controls (4.3,0.39) and (9.04,1.82) .. (14.21,4.28)   ;
\draw [color={rgb, 255:red, 230; green, 172; blue, 90 }  ,draw opacity=1 ][line width=1.5]    (496.31,204.09) -- (551.8,204.09) ;
\draw [shift={(554.8,204.09)}, rotate = 180] [color={rgb, 255:red, 230; green, 172; blue, 90 }  ,draw opacity=1 ][line width=1.5]    (14.21,-4.28) .. controls (9.04,-1.82) and (4.3,-0.39) .. (0,0) .. controls (4.3,0.39) and (9.04,1.82) .. (14.21,4.28)   ;
\draw [color={rgb, 255:red, 32; green, 148; blue, 131 }  ,draw opacity=0.5 ][line width=0.75]  [dash pattern={on 0.84pt off 2.51pt}]  (584.8,186.46) -- (593.82,175.87) ;
\draw [shift={(595.12,174.34)}, rotate = 490.4] [color={rgb, 255:red, 32; green, 148; blue, 131 }  ,draw opacity=0.5 ][line width=0.75]    (10.93,-3.29) .. controls (6.95,-1.4) and (3.31,-0.3) .. (0,0) .. controls (3.31,0.3) and (6.95,1.4) .. (10.93,3.29)   ;
\draw [color={rgb, 255:red, 32; green, 148; blue, 131 }  ,draw opacity=1 ][line width=2.25]  [dash pattern={on 2.53pt off 3.02pt}]  (496.31,189.73) -- (509.07,177.52) ;
\draw [shift={(511.96,174.75)}, rotate = 496.26] [color={rgb, 255:red, 32; green, 148; blue, 131 }  ,draw opacity=1 ][line width=2.25]    (17.49,-5.26) .. controls (11.12,-2.23) and (5.29,-0.48) .. (0,0) .. controls (5.29,0.48) and (11.12,2.23) .. (17.49,5.26)   ;
\draw [color={rgb, 255:red, 230; green, 172; blue, 90 }  ,draw opacity=0.5 ][line width=0.75]    (541.96,160.48) -- (593.12,160.76) ;
\draw [shift={(595.12,160.77)}, rotate = 180.31] [color={rgb, 255:red, 230; green, 172; blue, 90 }  ,draw opacity=0.5 ][line width=0.75]    (10.93,-3.29) .. controls (6.95,-1.4) and (3.31,-0.3) .. (0,0) .. controls (3.31,0.3) and (6.95,1.4) .. (10.93,3.29)   ;
\draw [color={rgb, 255:red, 213; green, 65; blue, 90 }  ,draw opacity=1 ][line width=1.5]    (527.65,84.11) -- (528.17,133.39)(524.65,84.14) -- (525.17,133.42) ;
\draw [shift={(526.77,142.4)}, rotate = 269.39] [color={rgb, 255:red, 213; green, 65; blue, 90 }  ,draw opacity=1 ][line width=1.5]    (14.21,-4.28) .. controls (9.04,-1.82) and (4.3,-0.39) .. (0,0) .. controls (4.3,0.39) and (9.04,1.82) .. (14.21,4.28)   ;
\draw [color={rgb, 255:red, 213; green, 65; blue, 90 }  ,draw opacity=0.5 ][line width=0.75]    (608.12,85.13) -- (608.12,136.33)(605.12,85.13) -- (605.12,136.33) ;
\draw [shift={(606.62,144.33)}, rotate = 270] [color={rgb, 255:red, 213; green, 65; blue, 90 }  ,draw opacity=0.5 ][line width=0.75]    (10.93,-3.29) .. controls (6.95,-1.4) and (3.31,-0.3) .. (0,0) .. controls (3.31,0.3) and (6.95,1.4) .. (10.93,3.29)   ;
\draw [color={rgb, 255:red, 213; green, 65; blue, 90 }  ,draw opacity=1 ][line width=1.5]    (572.1,131.25) -- (571.6,177.1)(569.1,131.22) -- (568.6,177.07) ;
\draw [shift={(570,186.09)}, rotate = 270.63] [color={rgb, 255:red, 213; green, 65; blue, 90 }  ,draw opacity=1 ][line width=1.5]    (14.21,-4.28) .. controls (9.04,-1.82) and (4.3,-0.39) .. (0,0) .. controls (4.3,0.39) and (9.04,1.82) .. (14.21,4.28)   ;

\end{tikzpicture}

\caption{2-polytope with a 0-polytope deletion and a 1-polytope addition.}
    \label{fig:mix_irregular_subexample}
\end{subfigure}
 \caption{3 irregular polytopes.}
 \label{fig:irregular_polytopes}
\end{figure}

3 irregular polytopes, respectively with deletion, addition and both, are shown in Figure~\ref{fig:irregular_polytopes}. These polytopes were introduced in~\cite{theseGuichaoua}, and more details are to be found in this work.

\subsection{System and Contrast (S\&C)}
\label{sec:s_and_c}
The core of the polytopic analysis lies in the fact that edges between elements model their relations. In that viewpoint, edges can link two elements which are not consecutive in the chronological order, and, hence, model non-sequential relations. This viewpoint is exploited in order to try to anticipate some relations.

Anticipation follows the ``System and Contrast'' (S\&C) model, developed by Bimbot and al.~\cite{bimbot2016system}. The S\&C model considers a passage of 4 elements, and, by studying the relations between the first 3 elements, tries to anticipate a ``fictive'' fourth element, compared with the real one.

Formally, when studying $f_{1/2}$, the relation between $a_1$ and $a_2$, and $f_{1/3}$, the relation between $a_1$ and $a_3$, the 3 elements are now represented by an element ($a_1$) and two relations ($f_{1/2}$ and $f_{1/3}$). Then, by composing $f_{1/2}$ and $f_{1/3}$, this model defines a fictive fourth element $\hat{a_4}$, implied by the first 3 elements, as $\hat{a_4} = f_{1/2}f_{1/3}.a_1$. The actual fourth element $a_4$ is then compared to this fictive one, which defines a ``contrast'' relation $\gamma$ as $\gamma_{\hat{4}/4}$.

When the fourth element $a_4$ is equal to the fictive fourth element, $a_4$ can be deduced from the first 3 elements, thus reducing the amount of information necessary to model 4 elements to 1 element (the first) and two relations. Otherwise, the fourth element is modeled with the contrast relation.

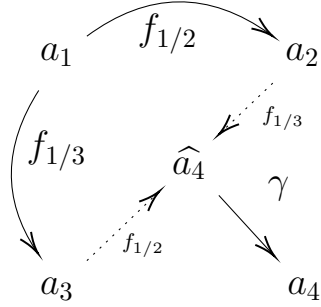
\begin{figure}[ht]
\centering
\begin{tikzpicture}[x=0.75pt,y=0.75pt,yscale=-1,xscale=1]

\draw (257.62,76.48) node  [font=\Large]  {$a_{1}$};
\draw (257.62,194.75) node  [font=\Large]  {$a_{3}$};
\draw (381.37,76.48) node  [font=\Large]  {$a_{2}$};
\draw (324.17,131.03) node  [font=\Large]  {$\widehat{a_{4}}$};
\draw (382.34,194.71) node  [font=\Large]  {$a_{4}$};
\draw (362.66,136.87) node [anchor=north west][inner sep=0.75pt]  [font=\Large] [align=left] {$\displaystyle \gamma $};
\draw (297.13,54.45) node [anchor=north west][inner sep=0.75pt]  [font=\Large] [align=left] {$\displaystyle f_{1/2}$};
\draw (289.13,165.45) node [anchor=north west][inner sep=0.75pt]  [font=\footnotesize] [align=left] {$\displaystyle f_{1/2}$};
\draw (359.77,101.27) node [anchor=north west][inner sep=0.75pt]  [font=\footnotesize] [align=left] {$\displaystyle f_{1/3}$};
\draw (241.13,113.45) node [anchor=north west][inner sep=0.75pt]  [font=\Large] [align=left] {$\displaystyle f_{1/3}$};
\draw    (272.62,67.1) .. controls (303.4,44.35) and (334.18,44.01) .. (364.96,66.08) ;
\draw [shift={(366.37,67.1)}, rotate = 216.48] [color={rgb, 255:red, 0; green, 0; blue, 0 }  ][line width=0.75]    (10.93,-3.29) .. controls (6.95,-1.4) and (3.31,-0.3) .. (0,0) .. controls (3.31,0.3) and (6.95,1.4) .. (10.93,3.29)   ;
\draw    (248.34,94.48) .. controls (230.42,122.54) and (229.92,149.43) .. (246.88,175.18) ;
\draw [shift={(247.94,176.75)}, rotate = 235.49] [color={rgb, 255:red, 0; green, 0; blue, 0 }  ][line width=0.75]    (10.93,-3.29) .. controls (6.95,-1.4) and (3.31,-0.3) .. (0,0) .. controls (3.31,0.3) and (6.95,1.4) .. (10.93,3.29)   ;
\draw  [dash pattern={on 0.84pt off 2.51pt}]  (272.62,180.39) -- (307.73,146.77) ;
\draw [shift={(309.17,145.39)}, rotate = 496.24] [color={rgb, 255:red, 0; green, 0; blue, 0 }  ][line width=0.75]    (10.93,-3.29) .. controls (6.95,-1.4) and (3.31,-0.3) .. (0,0) .. controls (3.31,0.3) and (6.95,1.4) .. (10.93,3.29)   ;
\draw    (339.17,147.45) -- (365.99,176.81) ;
\draw [shift={(367.34,178.29)}, rotate = 227.59] [color={rgb, 255:red, 0; green, 0; blue, 0 }  ][line width=0.75]    (10.93,-3.29) .. controls (6.95,-1.4) and (3.31,-0.3) .. (0,0) .. controls (3.31,0.3) and (6.95,1.4) .. (10.93,3.29)   ;
\draw  [dash pattern={on 0.84pt off 2.51pt}]  (366.37,90.79) -- (340.62,115.34) ;
\draw [shift={(339.17,116.72)}, rotate = 316.36] [color={rgb, 255:red, 0; green, 0; blue, 0 }  ][line width=0.75]    (10.93,-3.29) .. controls (6.95,-1.4) and (3.31,-0.3) .. (0,0) .. controls (3.31,0.3) and (6.95,1.4) .. (10.93,3.29)   ;

\end{tikzpicture}

  \caption{System \& Contrast model.}
  \label{fig:s_and_c}
\end{figure}

\section{Polytopic Complexity Cost of a Musical Passage}
\label{sec:guichaoua}
Now, starting with polytopes and with the S\&C model, this section defines a polytopic complexity cost $C(S,P)$ for a musical passage $S = \{a_1, a_2, ..., a_n\}$ on a polytope $P$. The polytopic cost is first defined as the sum of the individual costs of each element. Let us start with two useful definitions.

\begin{definition}[Primer] The first element in the polytope (and in the passage) is called ``primer''. The primer does not have any antecedent\footnote{The primer is in fact the only element without antecedent as deleting an element implies the deletion of its successors.}.
\end{definition}

\begin{definition}[Under-primer] Elements whose only antecedent is the primer are called ``under-primers''. Hence, under-primers are the only successors of the primer.
\end{definition}

\subsection{Information-Theory-like Viewpoint}
In polytopic analysis of music, an element is studied in comparison with previous elements (not necessarily consecutive in chronological order). The cost of an element $a_i$ is denoted as $C(a_i|\{a_1, a_2, ..., a_{i-1}\}, P)$.

This viewpoint aims at finding economical representations of a musical passage, where elements are only encoded if they can't be described by previous elements. This idea is close to the Minimum Description Length paradigm (MDL), an Information-Theory point of view where the shortest description (in terms of quantity of information) is considered the best one.

Here, elements are represented by their relations rather than being entirely described, and some of these relations are anticipated within the S\&C model.

All relations belonging to a same group $G_r$, they can be encoded by a same quantity of information $q$ (for example, representing the number of steps in the circle of triads between two elements leads to a set of 24 relations, requiring an encoding with $5$ bits to be entirely described). This quantity of information could be influenced by priors over the distribution of relations or by expert knowledge, but we do not explore that lead in this work. We can further simplify the model by considering that $q = 1$.

Concretely, this means that the complexity cost of a relation is 0 if the relation is the identity, or 1 otherwise. In addition, $a_1$ can't be described by previous elements, so $C(a_1 | P) = 1$. Finally,
\begin{equation}
    C(S,P) = C(a_1) + \sum_{i = 2}^{n} C(a_i|\{a_1, a_2, ..., a_{i-1}\}, P)
\end{equation}
with $C(a_i|\{a_1, a_2, ..., a_{i-1}\}, P) \in \{0,1\}, \forall 2 \leq i \leq n$.


\subsection{2-Polytope (Square, 4 elements)}
The core of the implication system lies on 4 elements polytopes (which are squares). A square polytope is composed of a primer $a_1$, two under-primers $a_2$ and $a_3$, and a fourth element $a_4$, which has both under-primers as antecedents. This polytope is evaluated as a S\&C model.

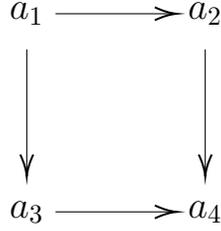
\begin{figure}[ht]
    \centering
    \begin{tikzpicture}[x=0.75pt,y=0.75pt,yscale=-1,xscale=1]

\draw (294,102) node  [font=\Large]  {$a_{1}$};
\draw (294,202) node  [font=\Large]  {$a_{3}$};
\draw (384,102) node  [font=\Large]  {$a_{2}$};
\draw (384,202) node  [font=\Large]  {$a_{4}$};
\draw    (308.5,102) -- (367.5,102) ;
\draw [shift={(369.5,102)}, rotate = 180] [color={rgb, 255:red, 0; green, 0; blue, 0 }  ][line width=0.75]    (10.93,-3.29) .. controls (6.95,-1.4) and (3.31,-0.3) .. (0,0) .. controls (3.31,0.3) and (6.95,1.4) .. (10.93,3.29)   ;
\draw    (294,120) -- (294,182) ;
\draw [shift={(294,184)}, rotate = 270] [color={rgb, 255:red, 0; green, 0; blue, 0 }  ][line width=0.75]    (10.93,-3.29) .. controls (6.95,-1.4) and (3.31,-0.3) .. (0,0) .. controls (3.31,0.3) and (6.95,1.4) .. (10.93,3.29)   ;
\draw    (308.5,202) -- (367.5,202) ;
\draw [shift={(369.5,202)}, rotate = 180] [color={rgb, 255:red, 0; green, 0; blue, 0 }  ][line width=0.75]    (10.93,-3.29) .. controls (6.95,-1.4) and (3.31,-0.3) .. (0,0) .. controls (3.31,0.3) and (6.95,1.4) .. (10.93,3.29)   ;
\draw    (384,120) -- (384,182) ;
\draw [shift={(384,184)}, rotate = 270] [color={rgb, 255:red, 0; green, 0; blue, 0 }  ][line width=0.75]    (10.93,-3.29) .. controls (6.95,-1.4) and (3.31,-0.3) .. (0,0) .. controls (3.31,0.3) and (6.95,1.4) .. (10.93,3.29)   ;

\end{tikzpicture}

    \caption{Square system.}
\end{figure}

The primer must be encoded (as initialization of the passage), so the initial cost of the polytope is 1.

Then, representing each under-primer in the S\&C model, for instance $a_2$, falls on one of these two cases:
\begin{itemize}
    \item $a_2 = a_1$: in that case, the relation $f_{1/2}$ is the identity, so $C(a_2 | \{a_1\}, P) = 0$.
    \item $a_2 \neq a_1$: it is necessary to represent the new element $a_2$ with $f_{1/2} \neq id$, so $C(a_2 | \{a_1\}, P) = 1$.
\end{itemize}

The same principle applies for $a_3$ with relation $f_{1/3}$.

Finally, $a_4$ is evaluated in comparison with the fictive element $\hat{a_4} = f_{1/2}f_{1/3}.a_1$:
\begin{itemize}
    \item If $\hat{a_4} = a_4$, the contrast is null. Hence, $a_4$ is encoded with the identity relation, yielding $C(a_4 | \{a_1, a_2, a_3\}, P) = 0$.
    \item If $\hat{a_4} \neq a_4$, the contrast relation needs to be encoded to model $a_4$, and $C(a_4 | \{a_1, a_2, a_3\}, P) = 1$.
\end{itemize}

Here, because $\hat{a_4} = f_{1/2}.a_3$, checking if the contrast is null is equivalent to checking if $f_{1/2}.a_3 == a_4$, or, differently written, if the relation between $a_1$ and $a_2$ is equal to the relation between $a_3$ and $a_4$, \textit{i.e.} $f_{1/2} == f_{3/4}$.

In this test, $a_4$ is evaluated via its antecedent $a_3$, and by comparing the relation $f_{3/4}$ with the parallel arrow starting from the primer ($f_{1/2}$). We define as \textbf{pivot element} the extremity of this parallel arrow. In this case, the pivot of $a_4$ related to its antecedent $a_3$ is the element $a_2$, denoted $p_4^3$.

\begin{definition}[Pivot] In general, we define the pivot of an element $a_i$ related to its antecedent $a_j$ the extremity of a relation parallel to $f_{j/i}$ and having the primer as origin. It is denoted $p_i^j$.
\end{definition}

By construction of polytopes, there always exists a pivot for elements which are not the primer or under-primers\footnote{It is obvious in square polytopes, and it can be generalized to every regular polytope. It is also possible to generalize to non-deleted and non-added vertices in irregular polytopes, because, by design, an element is necessarily deleted if one of its antecedent is deleted.}.

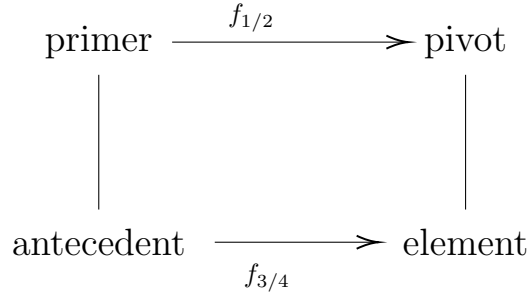
\begin{figure}[ht]
    \centering

\begin{tikzpicture}[x=0.75pt,y=0.75pt,yscale=-1,xscale=1]

\draw (294,101) node  [font=\Large]  {primer};
\draw (294,202) node  [font=\Large]  {antecedent};
\draw (479,101) node  [font=\Large]  {pivot};
\draw (479,202) node  [font=\Large]  {element};
\draw (370.83,88.75) node  [align=left] {\begin{minipage}[lt]{17.453356pt}\setlength\topsep{0pt}
$\displaystyle f_{1/2}$
\end{minipage}};
\draw (378.83,219.75) node  [align=left] {\begin{minipage}[lt]{17.453356pt}\setlength\topsep{0pt}
$\displaystyle f_{3/4}$
\end{minipage}};
\draw    (331,101) -- (448,101) ;
\draw [shift={(450,101)}, rotate = 180] [color={rgb, 255:red, 0; green, 0; blue, 0 }  ][line width=0.75]    (10.93,-3.29) .. controls (6.95,-1.4) and (3.31,-0.3) .. (0,0) .. controls (3.31,0.3) and (6.95,1.4) .. (10.93,3.29)   ;
\draw    (294,117.5) -- (294,185.5) ;
\draw    (352.5,202) -- (433,202) ;
\draw [shift={(435,202)}, rotate = 180] [color={rgb, 255:red, 0; green, 0; blue, 0 }  ][line width=0.75]    (10.93,-3.29) .. controls (6.95,-1.4) and (3.31,-0.3) .. (0,0) .. controls (3.31,0.3) and (6.95,1.4) .. (10.93,3.29)   ;
\draw    (479,117.5) -- (479,185.5) ;

\end{tikzpicture}

    \caption{Square polytope with antecedent and pivot.}
\end{figure}

\subsubsection{Equivalence of Both Couples Antecedent/Pivot for Square Systems}

It is important to notice that, thanks to the commutativity of the relation group, in a square polytope, the choice of the antecedent for $a_4$ is not important. Indeed, let's compare both cases:

\begin{itemize}
    \item Choosing $a_2$ as antecedent leads to choosing $a_3$ as pivot. Hence, testing the nullity of the contrast falls back to checking if $f_{1/3}.a_2 == a_4$. Yet, $a_2 = f_{1/2}.a_1$, so $f_{1/3}.a_2 = f_{1/3}f_{1/2}.a_1$, which leads to a test $f_{1/3}f_{1/2}.a_1 == a_4$.
    \item Choosing $a_3$ as antecedent leads to choosing $a_2$ as pivot.  Hence, testing the nullity of the contrast falls back to checking if $f_{1/2}.a_3 == a_4$. Yet, $a_3 = f_{1/3}.a_1$, so $f_{1/2}.a_3 = f_{1/2}f_{1/3}.a_1$, which leads to a test $f_{1/2}f_{1/3}.a_1 == a_4$.
\end{itemize}

With commutativity of relations, $f_{1/3}f_{1/2}.a_1 = f_{1/2}f_{1/3}.a_1$, so both tests are equivalents.

Finally, when, for two different antecedents of an element, one is the pivot of the other, \textbf{it is equivalent to choose either one as the antecedent and the other as pivot}. This is the case for all square polytopes.

Algorithm~\ref{alg:cost_square_system} sums up the previous rules, as a complexity cost function for a 4-elements musical passage on a square polytope.

\begin{algorithm}[ht]
\caption{Compression cost of a square polytope}
\begin{algorithmic}
\algnewcommand\algorithmicinput{\textbf{Input:}}
\algnewcommand\Input{\item[\algorithmicinput]}
\algnewcommand\algorithmicoutput{\textbf{Output:}}
\algnewcommand\Output{\item[\algorithmicoutput]}
\Input 4 elements $\{a_1, a_2, a_3, a_4\} \in A^4$, and the abelian group $G_r$.
\Output Cost $c$
\State c = 1 \Comment{Cost to encode $a_1$}
\For{$x = a_2, a_3$}
\If{$x == a_1$}
\State $c = c$ \Comment{Information is redundant.}
\Else
\State $c = c + 1$ \Comment{This relation must be encoded.}
\EndIf
\EndFor
\If{$f_{12} == f_{34}$}
\State $c = c$ \Comment{Information is redundant.}
\Else
\State $c = c + 1$ \Comment{Contrast needs to be encoded.}
\EndIf
\Return c
\end{algorithmic}
\label{alg:cost_square_system}
\end{algorithm}

\subsection{3-Polytope (Cube, 8 elements)}
Now, let's consider 3-polytopes, \textit{i.e.} cube polytopes, as presented in Figure \ref{fig:3_dim_polytope}. In this polytope, the primer is $a_1$, and the 3 sub-primers are $a_2$, $a_3$ and $a_5$.

\begin{figure}[ht]
    \centering
\begin{tikzpicture}[x=0.75pt,y=0.75pt,yscale=-1,xscale=1]

\draw [color={rgb, 255:red, 80; green, 227; blue, 194 }  ,draw opacity=1 ]   (53.09,142.99) .. controls (55.45,143.06) and (56.59,144.28) .. (56.52,146.63) .. controls (56.45,148.98) and (57.59,150.2) .. (59.94,150.28) .. controls (62.29,150.35) and (63.43,151.57) .. (63.36,153.92) .. controls (63.29,156.28) and (64.43,157.5) .. (66.79,157.57) .. controls (69.14,157.64) and (70.28,158.86) .. (70.21,161.21) .. controls (70.14,163.56) and (71.28,164.78) .. (73.63,164.86) .. controls (75.98,164.93) and (77.13,166.15) .. (77.06,168.5) .. controls (76.99,170.85) and (78.13,172.07) .. (80.48,172.14) .. controls (82.83,172.22) and (83.97,173.44) .. (83.9,175.79) .. controls (83.83,178.14) and (84.98,179.36) .. (87.33,179.43) .. controls (89.68,179.51) and (90.82,180.73) .. (90.75,183.08) .. controls (90.68,185.43) and (91.82,186.65) .. (94.17,186.72) .. controls (96.52,186.79) and (97.67,188.01) .. (97.6,190.36) .. controls (97.53,192.71) and (98.67,193.93) .. (101.02,194.01) .. controls (103.37,194.08) and (104.51,195.3) .. (104.44,197.65) -- (107.56,200.97) -- (109.62,203.16)(50.91,145.04) .. controls (53.26,145.12) and (54.4,146.34) .. (54.33,148.69) .. controls (54.26,151.04) and (55.4,152.26) .. (57.75,152.33) .. controls (60.11,152.4) and (61.25,153.62) .. (61.18,155.98) .. controls (61.11,158.33) and (62.25,159.55) .. (64.6,159.62) .. controls (66.95,159.69) and (68.09,160.91) .. (68.02,163.26) .. controls (67.95,165.62) and (69.09,166.84) .. (71.45,166.91) .. controls (73.8,166.98) and (74.94,168.2) .. (74.87,170.55) .. controls (74.8,172.9) and (75.94,174.12) .. (78.29,174.2) .. controls (80.64,174.27) and (81.79,175.49) .. (81.72,177.84) .. controls (81.65,180.19) and (82.79,181.41) .. (85.14,181.49) .. controls (87.49,181.56) and (88.63,182.78) .. (88.56,185.13) .. controls (88.49,187.48) and (89.64,188.7) .. (91.99,188.77) .. controls (94.34,188.85) and (95.48,190.07) .. (95.41,192.42) .. controls (95.34,194.77) and (96.48,195.99) .. (98.83,196.06) .. controls (101.19,196.13) and (102.33,197.35) .. (102.26,199.71) -- (105.38,203.03) -- (107.43,205.21) ;
\draw [shift={(114,210.02)}, rotate = 226.79] [color={rgb, 255:red, 80; green, 227; blue, 194 }  ,draw opacity=1 ][line width=0.75]    (10.93,-3.29) .. controls (6.95,-1.4) and (3.31,-0.3) .. (0,0) .. controls (3.31,0.3) and (6.95,1.4) .. (10.93,3.29)   ;
\draw [color={rgb, 255:red, 80; green, 227; blue, 194 }  ,draw opacity=1 ]   (109.14,82.04) .. controls (111.49,82.21) and (112.57,83.48) .. (112.4,85.83) .. controls (112.22,88.18) and (113.31,89.45) .. (115.66,89.62) .. controls (118.01,89.81) and (119.09,91.07) .. (118.91,93.42) .. controls (118.73,95.77) and (119.82,97.04) .. (122.17,97.21) .. controls (124.52,97.38) and (125.61,98.65) .. (125.43,101) .. controls (125.25,103.35) and (126.34,104.62) .. (128.69,104.79) .. controls (131.04,104.96) and (132.13,106.23) .. (131.95,108.58) .. controls (131.77,110.93) and (132.86,112.2) .. (135.21,112.38) .. controls (137.56,112.55) and (138.65,113.82) .. (138.47,116.17) .. controls (138.29,118.52) and (139.38,119.79) .. (141.73,119.96) .. controls (144.08,120.14) and (145.16,121.4) .. (144.98,123.75) .. controls (144.8,126.1) and (145.89,127.37) .. (148.24,127.54) .. controls (150.59,127.72) and (151.68,128.99) .. (151.5,131.34) .. controls (151.32,133.69) and (152.41,134.96) .. (154.76,135.13) -- (156.97,137.7) -- (158.92,139.97)(106.86,83.99) .. controls (109.21,84.17) and (110.3,85.44) .. (110.12,87.79) .. controls (109.94,90.14) and (111.03,91.41) .. (113.38,91.58) .. controls (115.73,91.75) and (116.82,93.02) .. (116.64,95.37) .. controls (116.46,97.72) and (117.55,98.99) .. (119.9,99.16) .. controls (122.25,99.33) and (123.34,100.6) .. (123.16,102.95) .. controls (122.98,105.3) and (124.07,106.57) .. (126.42,106.75) .. controls (128.77,106.93) and (129.85,108.19) .. (129.67,110.54) .. controls (129.49,112.89) and (130.58,114.16) .. (132.93,114.33) .. controls (135.28,114.5) and (136.37,115.77) .. (136.19,118.12) .. controls (136.01,120.47) and (137.1,121.74) .. (139.45,121.92) .. controls (141.8,122.09) and (142.89,123.36) .. (142.71,125.71) .. controls (142.53,128.06) and (143.62,129.33) .. (145.97,129.5) .. controls (148.32,129.67) and (149.41,130.94) .. (149.23,133.29) .. controls (149.05,135.64) and (150.14,136.91) .. (152.49,137.08) -- (154.69,139.65) -- (156.65,141.93) ;
\draw [shift={(163,147.02)}, rotate = 229.32999999999998] [color={rgb, 255:red, 80; green, 227; blue, 194 }  ,draw opacity=1 ][line width=0.75]    (10.93,-3.29) .. controls (6.95,-1.4) and (3.31,-0.3) .. (0,0) .. controls (3.31,0.3) and (6.95,1.4) .. (10.93,3.29)   ;
\draw [color={rgb, 255:red, 245; green, 166; blue, 35 }  ,draw opacity=1 ]   (250.06,132.29) .. controls (252.42,132.28) and (253.6,133.46) .. (253.6,135.82) .. controls (253.59,138.18) and (254.77,139.36) .. (257.13,139.36) .. controls (259.49,139.36) and (260.67,140.54) .. (260.67,142.9) -- (261.28,143.51) -- (263.4,145.63)(247.94,134.41) .. controls (250.29,134.41) and (251.47,135.59) .. (251.47,137.95) .. controls (251.47,140.31) and (252.65,141.49) .. (255.01,141.48) .. controls (257.37,141.48) and (258.55,142.66) .. (258.55,145.02) -- (259.16,145.63) -- (261.28,147.75) ;
\draw [shift={(268,152.35)}, rotate = 225] [color={rgb, 255:red, 245; green, 166; blue, 35 }  ,draw opacity=1 ][line width=0.75]    (10.93,-3.29) .. controls (6.95,-1.4) and (3.31,-0.3) .. (0,0) .. controls (3.31,0.3) and (6.95,1.4) .. (10.93,3.29)   ;
\draw [color={rgb, 255:red, 245; green, 166; blue, 35 }  ,draw opacity=1 ]   (340.99,136.22) .. controls (343.34,136.06) and (344.59,137.16) .. (344.75,139.51) .. controls (344.9,141.86) and (346.16,142.96) .. (348.51,142.81) -- (348.71,142.98) -- (350.97,144.95)(339.01,138.48) .. controls (341.36,138.32) and (342.62,139.42) .. (342.78,141.77) .. controls (342.93,144.12) and (344.19,145.22) .. (346.54,145.06) -- (346.73,145.24) -- (348.99,147.21) ;
\draw [shift={(356,151.35)}, rotate = 221.19] [color={rgb, 255:red, 245; green, 166; blue, 35 }  ,draw opacity=1 ][line width=0.75]    (10.93,-3.29) .. controls (6.95,-1.4) and (3.31,-0.3) .. (0,0) .. controls (3.31,0.3) and (6.95,1.4) .. (10.93,3.29)   ;
\draw [color={rgb, 255:red, 184; green, 233; blue, 134 }  ,draw opacity=1 ]   (461.38,194.65) .. controls (462.21,192.44) and (463.72,191.75) .. (465.93,192.58) .. controls (468.14,193.41) and (469.65,192.72) .. (470.48,190.51) .. controls (471.31,188.3) and (472.82,187.61) .. (475.03,188.43) .. controls (477.24,189.26) and (478.75,188.57) .. (479.58,186.36) .. controls (480.41,184.15) and (481.92,183.46) .. (484.13,184.29) .. controls (486.34,185.11) and (487.85,184.42) .. (488.68,182.21) .. controls (489.51,180) and (491.02,179.31) .. (493.23,180.14) .. controls (495.44,180.97) and (496.95,180.28) .. (497.78,178.07) .. controls (498.61,175.86) and (500.12,175.17) .. (502.33,176) .. controls (504.54,176.82) and (506.05,176.13) .. (506.88,173.92) .. controls (507.71,171.71) and (509.22,171.02) .. (511.43,171.85) .. controls (513.64,172.68) and (515.15,171.99) .. (515.98,169.78) .. controls (516.81,167.57) and (518.32,166.88) .. (520.53,167.7) .. controls (522.74,168.53) and (524.25,167.84) .. (525.08,165.63) .. controls (525.91,163.42) and (527.42,162.73) .. (529.63,163.56) .. controls (531.84,164.39) and (533.35,163.7) .. (534.18,161.49) .. controls (535.01,159.28) and (536.52,158.59) .. (538.73,159.41) -- (541.37,158.21) -- (544.1,156.97)(462.62,197.38) .. controls (463.45,195.17) and (464.96,194.48) .. (467.17,195.31) .. controls (469.38,196.14) and (470.89,195.45) .. (471.72,193.24) .. controls (472.55,191.03) and (474.06,190.34) .. (476.27,191.16) .. controls (478.48,191.99) and (479.99,191.3) .. (480.82,189.09) .. controls (481.65,186.88) and (483.16,186.19) .. (485.37,187.02) .. controls (487.58,187.84) and (489.09,187.15) .. (489.92,184.94) .. controls (490.75,182.73) and (492.26,182.04) .. (494.47,182.87) .. controls (496.68,183.7) and (498.19,183.01) .. (499.02,180.8) .. controls (499.85,178.59) and (501.36,177.9) .. (503.57,178.73) .. controls (505.78,179.55) and (507.29,178.86) .. (508.12,176.65) .. controls (508.95,174.44) and (510.46,173.75) .. (512.67,174.58) .. controls (514.88,175.41) and (516.39,174.72) .. (517.22,172.51) .. controls (518.05,170.3) and (519.56,169.61) .. (521.77,170.43) .. controls (523.98,171.26) and (525.49,170.57) .. (526.32,168.36) .. controls (527.15,166.15) and (528.66,165.46) .. (530.87,166.29) .. controls (533.08,167.12) and (534.59,166.43) .. (535.42,164.22) .. controls (536.25,162.01) and (537.76,161.32) .. (539.97,162.14) -- (542.61,160.94) -- (545.34,159.7) ;
\draw [shift={(552,155.02)}, rotate = 515.51] [color={rgb, 255:red, 184; green, 233; blue, 134 }  ,draw opacity=1 ][line width=0.75]    (10.93,-3.29) .. controls (6.95,-1.4) and (3.31,-0.3) .. (0,0) .. controls (3.31,0.3) and (6.95,1.4) .. (10.93,3.29)   ;
\draw [color={rgb, 255:red, 184; green, 233; blue, 134 }  ,draw opacity=1 ]   (461.43,100.63) .. controls (462.33,98.45) and (463.87,97.81) .. (466.05,98.72) .. controls (468.23,99.63) and (469.77,98.99) .. (470.67,96.81) .. controls (471.57,94.63) and (473.11,93.99) .. (475.29,94.9) .. controls (477.47,95.8) and (479.01,95.16) .. (479.91,92.98) .. controls (480.81,90.8) and (482.35,90.16) .. (484.53,91.07) .. controls (486.71,91.98) and (488.25,91.34) .. (489.15,89.16) .. controls (490.05,86.98) and (491.59,86.34) .. (493.77,87.25) .. controls (495.95,88.16) and (497.49,87.52) .. (498.39,85.34) .. controls (499.29,83.16) and (500.83,82.52) .. (503.01,83.42) .. controls (505.19,84.33) and (506.73,83.69) .. (507.63,81.51) .. controls (508.53,79.33) and (510.07,78.69) .. (512.25,79.6) .. controls (514.43,80.51) and (515.97,79.87) .. (516.87,77.69) .. controls (517.77,75.51) and (519.31,74.87) .. (521.49,75.78) .. controls (523.67,76.69) and (525.21,76.05) .. (526.11,73.87) .. controls (527.01,71.69) and (528.55,71.05) .. (530.73,71.95) .. controls (532.91,72.86) and (534.45,72.22) .. (535.35,70.04) -- (538.26,68.84) -- (541.03,67.69)(462.57,103.4) .. controls (463.47,101.23) and (465.01,100.59) .. (467.19,101.49) .. controls (469.37,102.4) and (470.91,101.76) .. (471.81,99.58) .. controls (472.71,97.4) and (474.25,96.76) .. (476.43,97.67) .. controls (478.61,98.58) and (480.15,97.94) .. (481.05,95.76) .. controls (481.95,93.58) and (483.49,92.94) .. (485.67,93.84) .. controls (487.85,94.75) and (489.39,94.11) .. (490.29,91.93) .. controls (491.19,89.75) and (492.73,89.11) .. (494.91,90.02) .. controls (497.09,90.93) and (498.63,90.29) .. (499.53,88.11) .. controls (500.43,85.93) and (501.97,85.29) .. (504.15,86.2) .. controls (506.33,87.11) and (507.87,86.47) .. (508.77,84.29) .. controls (509.67,82.11) and (511.21,81.47) .. (513.39,82.37) .. controls (515.57,83.28) and (517.11,82.64) .. (518.01,80.46) .. controls (518.91,78.28) and (520.45,77.64) .. (522.63,78.55) .. controls (524.81,79.46) and (526.35,78.82) .. (527.25,76.64) .. controls (528.15,74.46) and (529.69,73.82) .. (531.87,74.73) .. controls (534.05,75.63) and (535.59,74.99) .. (536.49,72.81) -- (539.41,71.61) -- (542.18,70.46) ;
\draw [shift={(549,66.02)}, rotate = 517.52] [color={rgb, 255:red, 184; green, 233; blue, 134 }  ,draw opacity=1 ][line width=0.75]    (10.93,-3.29) .. controls (6.95,-1.4) and (3.31,-0.3) .. (0,0) .. controls (3.31,0.3) and (6.95,1.4) .. (10.93,3.29)   ;

\draw (40,119) node  [font=\Large]  {$a_{1}$};
\draw (90,68) node  [font=\Large]  {$a_{5}$};
\draw (40,219) node  [font=\Large]  {$a_{3}$};
\draw (170,67) node  [font=\Large]  {$a_{6}$};
\draw (130,120) node  [font=\Large]  {$a_{2}$};
\draw (90,159) node  [font=\Large]  {$a_{7}$};
\draw (170,159) node  [font=\Large]  {$a_{8}$};
\draw (130,219) node  [font=\Large]  {$a_{4}$};
\draw (233,119) node  [font=\Large]  {$a_{1}$};
\draw (283,68) node  [font=\Large]  {$a_{5}$};
\draw (233,219) node  [font=\Large]  {$a_{3}$};
\draw (363,67) node  [font=\Large]  {$a_{6}$};
\draw (323,120) node  [font=\Large]  {$a_{2}$};
\draw (283,159) node  [font=\Large]  {$a_{7}$};
\draw (363,159) node  [font=\Large]  {$a_{8}$};
\draw (323,219) node  [font=\Large]  {$a_{4}$};
\draw (442,107) node  [font=\Large]  {$a_{1}$};
\draw (492,56) node  [font=\Large]  {$a_{5}$};
\draw (442,207) node  [font=\Large]  {$a_{3}$};
\draw (572,55) node  [font=\Large]  {$a_{6}$};
\draw (532,108) node  [font=\Large]  {$a_{2}$};
\draw (492,147) node  [font=\Large]  {$a_{7}$};
\draw (572,147) node  [font=\Large]  {$a_{8}$};
\draw (532,207) node  [font=\Large]  {$a_{4}$};
\draw [color={rgb, 255:red, 0; green, 0; blue, 0 }  ,draw opacity=1 ] [dash pattern={on 0.84pt off 2.51pt}]  (104.5,67.82) -- (153.5,67.21) ;
\draw [shift={(155.5,67.18)}, rotate = 539.28] [color={rgb, 255:red, 0; green, 0; blue, 0 }  ,draw opacity=1 ][line width=0.75]    (10.93,-3.29) .. controls (6.95,-1.4) and (3.31,-0.3) .. (0,0) .. controls (3.31,0.3) and (6.95,1.4) .. (10.93,3.29)   ;
\draw [color={rgb, 255:red, 80; green, 227; blue, 194 }  ,draw opacity=1 ]   (53.43,103.16) .. controls (53.4,100.8) and (54.57,99.61) .. (56.93,99.59) .. controls (59.29,99.57) and (60.46,98.38) .. (60.43,96.02) .. controls (60.4,93.66) and (61.57,92.47) .. (63.93,92.45) -- (66.73,89.59) -- (68.83,87.45)(55.57,105.26) .. controls (55.55,102.91) and (56.72,101.72) .. (59.07,101.69) .. controls (61.43,101.67) and (62.6,100.48) .. (62.57,98.12) .. controls (62.54,95.76) and (63.71,94.57) .. (66.07,94.55) -- (68.87,91.69) -- (70.97,89.55) ;
\draw [shift={(75.5,82.79)}, rotate = 494.43] [color={rgb, 255:red, 80; green, 227; blue, 194 }  ,draw opacity=1 ][line width=0.75]    (10.93,-3.29) .. controls (6.95,-1.4) and (3.31,-0.3) .. (0,0) .. controls (3.31,0.3) and (6.95,1.4) .. (10.93,3.29)   ;
\draw [color={rgb, 255:red, 0; green, 0; blue, 0 }  ,draw opacity=1 ] [dash pattern={on 0.84pt off 2.51pt}]  (54.5,119.16) -- (113.5,119.82) ;
\draw [shift={(115.5,119.84)}, rotate = 180.64] [color={rgb, 255:red, 0; green, 0; blue, 0 }  ,draw opacity=1 ][line width=0.75]    (10.93,-3.29) .. controls (6.95,-1.4) and (3.31,-0.3) .. (0,0) .. controls (3.31,0.3) and (6.95,1.4) .. (10.93,3.29)   ;
\draw  [dash pattern={on 0.84pt off 2.51pt}]  (143.58,102) -- (155.21,86.6) ;
\draw [shift={(156.42,85)}, rotate = 487.04] [color={rgb, 255:red, 0; green, 0; blue, 0 }  ][line width=0.75]    (10.93,-3.29) .. controls (6.95,-1.4) and (3.31,-0.3) .. (0,0) .. controls (3.31,0.3) and (6.95,1.4) .. (10.93,3.29)   ;
\draw [color={rgb, 255:red, 0; green, 0; blue, 0 }  ,draw opacity=1 ] [dash pattern={on 0.84pt off 2.51pt}]  (40,137) -- (40,199) ;
\draw [shift={(40,201)}, rotate = 270] [color={rgb, 255:red, 0; green, 0; blue, 0 }  ,draw opacity=1 ][line width=0.75]    (10.93,-3.29) .. controls (6.95,-1.4) and (3.31,-0.3) .. (0,0) .. controls (3.31,0.3) and (6.95,1.4) .. (10.93,3.29)   ;
\draw [color={rgb, 255:red, 0; green, 0; blue, 0 }  ,draw opacity=1 ] [dash pattern={on 0.84pt off 2.51pt}]  (54.5,219) -- (113.5,219) ;
\draw [shift={(115.5,219)}, rotate = 180] [color={rgb, 255:red, 0; green, 0; blue, 0 }  ,draw opacity=1 ][line width=0.75]    (10.93,-3.29) .. controls (6.95,-1.4) and (3.31,-0.3) .. (0,0) .. controls (3.31,0.3) and (6.95,1.4) .. (10.93,3.29)   ;
\draw [color={rgb, 255:red, 80; green, 227; blue, 194 }  ,draw opacity=1 ]   (140.75,200.17) .. controls (140.29,197.86) and (141.22,196.47) .. (143.53,196.01) .. controls (145.84,195.55) and (146.77,194.16) .. (146.3,191.85) .. controls (145.83,189.54) and (146.76,188.15) .. (149.07,187.69) -- (150.65,185.32) -- (152.31,182.82)(143.25,201.83) .. controls (142.78,199.52) and (143.71,198.14) .. (146.02,197.67) .. controls (148.33,197.21) and (149.26,195.82) .. (148.8,193.51) .. controls (148.33,191.2) and (149.26,189.81) .. (151.57,189.35) -- (153.15,186.98) -- (154.81,184.49) ;
\draw [shift={(158,177)}, rotate = 483.69] [color={rgb, 255:red, 80; green, 227; blue, 194 }  ,draw opacity=1 ][line width=0.75]    (10.93,-3.29) .. controls (6.95,-1.4) and (3.31,-0.3) .. (0,0) .. controls (3.31,0.3) and (6.95,1.4) .. (10.93,3.29)   ;
\draw  [dash pattern={on 0.84pt off 2.51pt}]  (54.5,201.6) -- (74.22,177.94) ;
\draw [shift={(75.5,176.4)}, rotate = 489.81] [color={rgb, 255:red, 0; green, 0; blue, 0 }  ][line width=0.75]    (10.93,-3.29) .. controls (6.95,-1.4) and (3.31,-0.3) .. (0,0) .. controls (3.31,0.3) and (6.95,1.4) .. (10.93,3.29)   ;
\draw [color={rgb, 255:red, 0; green, 0; blue, 0 }  ,draw opacity=1 ] [dash pattern={on 0.84pt off 2.51pt}]  (104.5,159) -- (122.55,159)(130.55,159) -- (153.5,159) ;
\draw [shift={(155.5,159)}, rotate = 180] [color={rgb, 255:red, 0; green, 0; blue, 0 }  ,draw opacity=1 ][line width=0.75]    (10.93,-3.29) .. controls (6.95,-1.4) and (3.31,-0.3) .. (0,0) .. controls (3.31,0.3) and (6.95,1.4) .. (10.93,3.29)   ;
\draw [color={rgb, 255:red, 0; green, 0; blue, 0 }  ,draw opacity=1 ] [dash pattern={on 0.84pt off 2.51pt}]  (90,86) -- (90,111.68)(90,119.68) -- (90,139) ;
\draw [shift={(90,141)}, rotate = 270] [color={rgb, 255:red, 0; green, 0; blue, 0 }  ,draw opacity=1 ][line width=0.75]    (10.93,-3.29) .. controls (6.95,-1.4) and (3.31,-0.3) .. (0,0) .. controls (3.31,0.3) and (6.95,1.4) .. (10.93,3.29)   ;
\draw [color={rgb, 255:red, 0; green, 0; blue, 0 }  ,draw opacity=1 ] [dash pattern={on 0.84pt off 2.51pt}]  (170,85) -- (170,139) ;
\draw [shift={(170,141)}, rotate = 270] [color={rgb, 255:red, 0; green, 0; blue, 0 }  ,draw opacity=1 ][line width=0.75]    (10.93,-3.29) .. controls (6.95,-1.4) and (3.31,-0.3) .. (0,0) .. controls (3.31,0.3) and (6.95,1.4) .. (10.93,3.29)   ;
\draw [color={rgb, 255:red, 0; green, 0; blue, 0 }  ,draw opacity=1 ] [dash pattern={on 0.84pt off 2.51pt}]  (130,138) -- (130,199) ;
\draw [shift={(130,201)}, rotate = 270] [color={rgb, 255:red, 0; green, 0; blue, 0 }  ,draw opacity=1 ][line width=0.75]    (10.93,-3.29) .. controls (6.95,-1.4) and (3.31,-0.3) .. (0,0) .. controls (3.31,0.3) and (6.95,1.4) .. (10.93,3.29)   ;
\draw [color={rgb, 255:red, 0; green, 0; blue, 0 }  ,draw opacity=1 ] [dash pattern={on 0.84pt off 2.51pt}]  (297.5,67.82) -- (346.5,67.21) ;
\draw [shift={(348.5,67.18)}, rotate = 539.28] [color={rgb, 255:red, 0; green, 0; blue, 0 }  ,draw opacity=1 ][line width=0.75]    (10.93,-3.29) .. controls (6.95,-1.4) and (3.31,-0.3) .. (0,0) .. controls (3.31,0.3) and (6.95,1.4) .. (10.93,3.29)   ;
\draw [color={rgb, 255:red, 0; green, 0; blue, 0 }  ,draw opacity=1 ] [dash pattern={on 0.84pt off 2.51pt}]  (247.5,104.21) -- (267.1,84.22) ;
\draw [shift={(268.5,82.79)}, rotate = 494.43] [color={rgb, 255:red, 0; green, 0; blue, 0 }  ,draw opacity=1 ][line width=0.75]    (10.93,-3.29) .. controls (6.95,-1.4) and (3.31,-0.3) .. (0,0) .. controls (3.31,0.3) and (6.95,1.4) .. (10.93,3.29)   ;
\draw [color={rgb, 255:red, 245; green, 166; blue, 35 }  ,draw opacity=1 ]   (247.52,117.66) .. controls (249.21,116.01) and (250.87,116.03) .. (252.52,117.72) .. controls (254.17,119.4) and (255.84,119.42) .. (257.52,117.77) .. controls (259.21,116.12) and (260.87,116.14) .. (262.52,117.83) .. controls (264.17,119.51) and (265.84,119.53) .. (267.52,117.88) .. controls (269.21,116.23) and (270.87,116.25) .. (272.52,117.94) .. controls (274.17,119.62) and (275.83,119.64) .. (277.51,117.99) .. controls (279.2,116.34) and (280.86,116.36) .. (282.51,118.05) .. controls (284.16,119.74) and (285.82,119.76) .. (287.51,118.11) .. controls (289.19,116.46) and (290.86,116.48) .. (292.51,118.16) .. controls (294.16,119.85) and (295.82,119.87) .. (297.51,118.22) -- (297.52,118.22) -- (300.52,118.25)(247.48,120.66) .. controls (249.17,119.01) and (250.83,119.03) .. (252.48,120.72) .. controls (254.13,122.4) and (255.8,122.42) .. (257.48,120.77) .. controls (259.17,119.12) and (260.83,119.14) .. (262.48,120.83) .. controls (264.13,122.51) and (265.8,122.53) .. (267.48,120.88) .. controls (269.17,119.23) and (270.83,119.25) .. (272.48,120.94) .. controls (274.13,122.62) and (275.8,122.64) .. (277.48,120.99) .. controls (279.17,119.34) and (280.83,119.36) .. (282.48,121.05) .. controls (284.13,122.74) and (285.79,122.76) .. (287.48,121.11) .. controls (289.16,119.46) and (290.83,119.48) .. (292.48,121.16) .. controls (294.13,122.85) and (295.79,122.87) .. (297.48,121.22) -- (297.48,121.22) -- (300.48,121.25) ;
\draw [shift={(308.5,119.84)}, rotate = 180.64] [color={rgb, 255:red, 245; green, 166; blue, 35 }  ,draw opacity=1 ][line width=0.75]    (10.93,-3.29) .. controls (6.95,-1.4) and (3.31,-0.3) .. (0,0) .. controls (3.31,0.3) and (6.95,1.4) .. (10.93,3.29)   ;
\draw  [dash pattern={on 0.84pt off 2.51pt}]  (336.58,102) -- (348.21,86.6) ;
\draw [shift={(349.42,85)}, rotate = 487.04] [color={rgb, 255:red, 0; green, 0; blue, 0 }  ][line width=0.75]    (10.93,-3.29) .. controls (6.95,-1.4) and (3.31,-0.3) .. (0,0) .. controls (3.31,0.3) and (6.95,1.4) .. (10.93,3.29)   ;
\draw [color={rgb, 255:red, 0; green, 0; blue, 0 }  ,draw opacity=1 ] [dash pattern={on 0.84pt off 2.51pt}]  (233,137) -- (233,199) ;
\draw [shift={(233,201)}, rotate = 270] [color={rgb, 255:red, 0; green, 0; blue, 0 }  ,draw opacity=1 ][line width=0.75]    (10.93,-3.29) .. controls (6.95,-1.4) and (3.31,-0.3) .. (0,0) .. controls (3.31,0.3) and (6.95,1.4) .. (10.93,3.29)   ;
\draw [color={rgb, 255:red, 0; green, 0; blue, 0 }  ,draw opacity=1 ] [dash pattern={on 0.84pt off 2.51pt}]  (247.5,219) -- (306.5,219) ;
\draw [shift={(308.5,219)}, rotate = 180] [color={rgb, 255:red, 0; green, 0; blue, 0 }  ,draw opacity=1 ][line width=0.75]    (10.93,-3.29) .. controls (6.95,-1.4) and (3.31,-0.3) .. (0,0) .. controls (3.31,0.3) and (6.95,1.4) .. (10.93,3.29)   ;
\draw [color={rgb, 255:red, 0; green, 0; blue, 0 }  ,draw opacity=1 ] [dash pattern={on 0.84pt off 2.51pt}]  (335,201) -- (349.89,178.66) ;
\draw [shift={(351,177)}, rotate = 483.69] [color={rgb, 255:red, 0; green, 0; blue, 0 }  ,draw opacity=1 ][line width=0.75]    (10.93,-3.29) .. controls (6.95,-1.4) and (3.31,-0.3) .. (0,0) .. controls (3.31,0.3) and (6.95,1.4) .. (10.93,3.29)   ;
\draw  [dash pattern={on 0.84pt off 2.51pt}]  (247.5,201.6) -- (267.22,177.94) ;
\draw [shift={(268.5,176.4)}, rotate = 489.81] [color={rgb, 255:red, 0; green, 0; blue, 0 }  ][line width=0.75]    (10.93,-3.29) .. controls (6.95,-1.4) and (3.31,-0.3) .. (0,0) .. controls (3.31,0.3) and (6.95,1.4) .. (10.93,3.29)   ;
\draw [color={rgb, 255:red, 245; green, 166; blue, 35 }  ,draw opacity=1 ]   (297.5,157.5) .. controls (299.17,155.83) and (300.83,155.83) .. (302.5,157.5) .. controls (304.17,159.17) and (305.83,159.17) .. (307.5,157.5) .. controls (309.17,155.83) and (310.83,155.83) .. (312.5,157.5) .. controls (312.62,157.62) and (312.73,157.73) .. (312.85,157.83)(320.85,156.39) .. controls (321.4,156.58) and (321.95,156.95) .. (322.5,157.5) .. controls (324.17,159.17) and (325.83,159.17) .. (327.5,157.5) .. controls (329.17,155.83) and (330.83,155.83) .. (332.5,157.5) .. controls (334.17,159.17) and (335.83,159.17) .. (337.5,157.5) -- (337.5,157.5) -- (340.5,157.5)(297.5,160.5) .. controls (299.17,158.83) and (300.83,158.83) .. (302.5,160.5) .. controls (304.17,162.17) and (305.83,162.17) .. (307.5,160.5) .. controls (309.17,158.83) and (310.83,158.83) .. (312.5,160.5) .. controls (312.62,160.62) and (312.73,160.73) .. (312.85,160.83)(320.85,159.39) .. controls (321.4,159.58) and (321.95,159.95) .. (322.5,160.5) .. controls (324.17,162.17) and (325.83,162.17) .. (327.5,160.5) .. controls (329.17,158.83) and (330.83,158.83) .. (332.5,160.5) .. controls (334.17,162.17) and (335.83,162.17) .. (337.5,160.5) -- (337.5,160.5) -- (340.5,160.5) ;
\draw [shift={(348.5,159)}, rotate = 180] [color={rgb, 255:red, 245; green, 166; blue, 35 }  ,draw opacity=1 ][line width=0.75]    (10.93,-3.29) .. controls (6.95,-1.4) and (3.31,-0.3) .. (0,0) .. controls (3.31,0.3) and (6.95,1.4) .. (10.93,3.29)   ;
\draw [color={rgb, 255:red, 0; green, 0; blue, 0 }  ,draw opacity=1 ] [dash pattern={on 0.84pt off 2.51pt}]  (283,86) -- (283,111.68)(283,119.68) -- (283,139) ;
\draw [shift={(283,141)}, rotate = 270] [color={rgb, 255:red, 0; green, 0; blue, 0 }  ,draw opacity=1 ][line width=0.75]    (10.93,-3.29) .. controls (6.95,-1.4) and (3.31,-0.3) .. (0,0) .. controls (3.31,0.3) and (6.95,1.4) .. (10.93,3.29)   ;
\draw [color={rgb, 255:red, 0; green, 0; blue, 0 }  ,draw opacity=1 ] [dash pattern={on 0.84pt off 2.51pt}]  (363,85) -- (363,139) ;
\draw [shift={(363,141)}, rotate = 270] [color={rgb, 255:red, 0; green, 0; blue, 0 }  ,draw opacity=1 ][line width=0.75]    (10.93,-3.29) .. controls (6.95,-1.4) and (3.31,-0.3) .. (0,0) .. controls (3.31,0.3) and (6.95,1.4) .. (10.93,3.29)   ;
\draw [color={rgb, 255:red, 0; green, 0; blue, 0 }  ,draw opacity=1 ] [dash pattern={on 0.84pt off 2.51pt}]  (323,138) -- (323,199) ;
\draw [shift={(323,201)}, rotate = 270] [color={rgb, 255:red, 0; green, 0; blue, 0 }  ,draw opacity=1 ][line width=0.75]    (10.93,-3.29) .. controls (6.95,-1.4) and (3.31,-0.3) .. (0,0) .. controls (3.31,0.3) and (6.95,1.4) .. (10.93,3.29)   ;
\draw [color={rgb, 255:red, 0; green, 0; blue, 0 }  ,draw opacity=1 ] [dash pattern={on 0.84pt off 2.51pt}]  (506.5,55.82) -- (555.5,55.21) ;
\draw [shift={(557.5,55.18)}, rotate = 539.28] [color={rgb, 255:red, 0; green, 0; blue, 0 }  ,draw opacity=1 ][line width=0.75]    (10.93,-3.29) .. controls (6.95,-1.4) and (3.31,-0.3) .. (0,0) .. controls (3.31,0.3) and (6.95,1.4) .. (10.93,3.29)   ;
\draw [color={rgb, 255:red, 0; green, 0; blue, 0 }  ,draw opacity=1 ] [dash pattern={on 0.84pt off 2.51pt}]  (456.5,92.21) -- (476.1,72.22) ;
\draw [shift={(477.5,70.79)}, rotate = 494.43] [color={rgb, 255:red, 0; green, 0; blue, 0 }  ,draw opacity=1 ][line width=0.75]    (10.93,-3.29) .. controls (6.95,-1.4) and (3.31,-0.3) .. (0,0) .. controls (3.31,0.3) and (6.95,1.4) .. (10.93,3.29)   ;
\draw [color={rgb, 255:red, 0; green, 0; blue, 0 }  ,draw opacity=1 ] [dash pattern={on 0.84pt off 2.51pt}]  (456.5,107.16) -- (515.5,107.82) ;
\draw [shift={(517.5,107.84)}, rotate = 180.64] [color={rgb, 255:red, 0; green, 0; blue, 0 }  ,draw opacity=1 ][line width=0.75]    (10.93,-3.29) .. controls (6.95,-1.4) and (3.31,-0.3) .. (0,0) .. controls (3.31,0.3) and (6.95,1.4) .. (10.93,3.29)   ;
\draw  [dash pattern={on 0.84pt off 2.51pt}]  (545.58,90) -- (557.21,74.6) ;
\draw [shift={(558.42,73)}, rotate = 487.04] [color={rgb, 255:red, 0; green, 0; blue, 0 }  ][line width=0.75]    (10.93,-3.29) .. controls (6.95,-1.4) and (3.31,-0.3) .. (0,0) .. controls (3.31,0.3) and (6.95,1.4) .. (10.93,3.29)   ;
\draw [color={rgb, 255:red, 184; green, 233; blue, 134 }  ,draw opacity=1 ]   (443.5,125) .. controls (445.17,126.67) and (445.17,128.33) .. (443.5,130) .. controls (441.83,131.67) and (441.83,133.33) .. (443.5,135) .. controls (445.17,136.67) and (445.17,138.33) .. (443.5,140) .. controls (441.83,141.67) and (441.83,143.33) .. (443.5,145) .. controls (445.17,146.67) and (445.17,148.33) .. (443.5,150) .. controls (441.83,151.67) and (441.83,153.33) .. (443.5,155) .. controls (445.17,156.67) and (445.17,158.33) .. (443.5,160) .. controls (441.83,161.67) and (441.83,163.33) .. (443.5,165) .. controls (445.17,166.67) and (445.17,168.33) .. (443.5,170) .. controls (441.83,171.67) and (441.83,173.33) .. (443.5,175) -- (443.5,178) -- (443.5,181)(440.5,125) .. controls (442.17,126.67) and (442.17,128.33) .. (440.5,130) .. controls (438.83,131.67) and (438.83,133.33) .. (440.5,135) .. controls (442.17,136.67) and (442.17,138.33) .. (440.5,140) .. controls (438.83,141.67) and (438.83,143.33) .. (440.5,145) .. controls (442.17,146.67) and (442.17,148.33) .. (440.5,150) .. controls (438.83,151.67) and (438.83,153.33) .. (440.5,155) .. controls (442.17,156.67) and (442.17,158.33) .. (440.5,160) .. controls (438.83,161.67) and (438.83,163.33) .. (440.5,165) .. controls (442.17,166.67) and (442.17,168.33) .. (440.5,170) .. controls (438.83,171.67) and (438.83,173.33) .. (440.5,175) -- (440.5,178) -- (440.5,181) ;
\draw [shift={(442,189)}, rotate = 270] [color={rgb, 255:red, 184; green, 233; blue, 134 }  ,draw opacity=1 ][line width=0.75]    (10.93,-3.29) .. controls (6.95,-1.4) and (3.31,-0.3) .. (0,0) .. controls (3.31,0.3) and (6.95,1.4) .. (10.93,3.29)   ;
\draw [color={rgb, 255:red, 0; green, 0; blue, 0 }  ,draw opacity=1 ] [dash pattern={on 0.84pt off 2.51pt}]  (456.5,207) -- (515.5,207) ;
\draw [shift={(517.5,207)}, rotate = 180] [color={rgb, 255:red, 0; green, 0; blue, 0 }  ,draw opacity=1 ][line width=0.75]    (10.93,-3.29) .. controls (6.95,-1.4) and (3.31,-0.3) .. (0,0) .. controls (3.31,0.3) and (6.95,1.4) .. (10.93,3.29)   ;
\draw [color={rgb, 255:red, 0; green, 0; blue, 0 }  ,draw opacity=1 ] [dash pattern={on 0.84pt off 2.51pt}]  (544,189) -- (558.89,166.66) ;
\draw [shift={(560,165)}, rotate = 483.69] [color={rgb, 255:red, 0; green, 0; blue, 0 }  ,draw opacity=1 ][line width=0.75]    (10.93,-3.29) .. controls (6.95,-1.4) and (3.31,-0.3) .. (0,0) .. controls (3.31,0.3) and (6.95,1.4) .. (10.93,3.29)   ;
\draw  [dash pattern={on 0.84pt off 2.51pt}]  (456.5,189.6) -- (476.22,165.94) ;
\draw [shift={(477.5,164.4)}, rotate = 489.81] [color={rgb, 255:red, 0; green, 0; blue, 0 }  ][line width=0.75]    (10.93,-3.29) .. controls (6.95,-1.4) and (3.31,-0.3) .. (0,0) .. controls (3.31,0.3) and (6.95,1.4) .. (10.93,3.29)   ;
\draw [color={rgb, 255:red, 0; green, 0; blue, 0 }  ,draw opacity=1 ] [dash pattern={on 0.84pt off 2.51pt}]  (506.5,147) -- (524.55,147)(532.55,147) -- (555.5,147) ;
\draw [shift={(557.5,147)}, rotate = 180] [color={rgb, 255:red, 0; green, 0; blue, 0 }  ,draw opacity=1 ][line width=0.75]    (10.93,-3.29) .. controls (6.95,-1.4) and (3.31,-0.3) .. (0,0) .. controls (3.31,0.3) and (6.95,1.4) .. (10.93,3.29)   ;
\draw [color={rgb, 255:red, 0; green, 0; blue, 0 }  ,draw opacity=1 ] [dash pattern={on 0.84pt off 2.51pt}]  (492,74) -- (492,99.68)(492,107.68) -- (492,127) ;
\draw [shift={(492,129)}, rotate = 270] [color={rgb, 255:red, 0; green, 0; blue, 0 }  ,draw opacity=1 ][line width=0.75]    (10.93,-3.29) .. controls (6.95,-1.4) and (3.31,-0.3) .. (0,0) .. controls (3.31,0.3) and (6.95,1.4) .. (10.93,3.29)   ;
\draw [color={rgb, 255:red, 184; green, 233; blue, 134 }  ,draw opacity=1 ]   (573.5,73) .. controls (575.17,74.67) and (575.17,76.33) .. (573.5,78) .. controls (571.83,79.67) and (571.83,81.33) .. (573.5,83) .. controls (575.17,84.67) and (575.17,86.33) .. (573.5,88) .. controls (571.83,89.67) and (571.83,91.33) .. (573.5,93) .. controls (575.17,94.67) and (575.17,96.33) .. (573.5,98) .. controls (571.83,99.67) and (571.83,101.33) .. (573.5,103) .. controls (575.17,104.67) and (575.17,106.33) .. (573.5,108) .. controls (571.83,109.67) and (571.83,111.33) .. (573.5,113) .. controls (575.17,114.67) and (575.17,116.33) .. (573.5,118) -- (573.5,121)(570.5,73) .. controls (572.17,74.67) and (572.17,76.33) .. (570.5,78) .. controls (568.83,79.67) and (568.83,81.33) .. (570.5,83) .. controls (572.17,84.67) and (572.17,86.33) .. (570.5,88) .. controls (568.83,89.67) and (568.83,91.33) .. (570.5,93) .. controls (572.17,94.67) and (572.17,96.33) .. (570.5,98) .. controls (568.83,99.67) and (568.83,101.33) .. (570.5,103) .. controls (572.17,104.67) and (572.17,106.33) .. (570.5,108) .. controls (568.83,109.67) and (568.83,111.33) .. (570.5,113) .. controls (572.17,114.67) and (572.17,116.33) .. (570.5,118) -- (570.5,121) ;
\draw [shift={(572,129)}, rotate = 270] [color={rgb, 255:red, 184; green, 233; blue, 134 }  ,draw opacity=1 ][line width=0.75]    (10.93,-3.29) .. controls (6.95,-1.4) and (3.31,-0.3) .. (0,0) .. controls (3.31,0.3) and (6.95,1.4) .. (10.93,3.29)   ;
\draw [color={rgb, 255:red, 0; green, 0; blue, 0 }  ,draw opacity=1 ] [dash pattern={on 0.84pt off 2.51pt}]  (532,126) -- (532,187) ;
\draw [shift={(532,189)}, rotate = 270] [color={rgb, 255:red, 0; green, 0; blue, 0 }  ,draw opacity=1 ][line width=0.75]    (10.93,-3.29) .. controls (6.95,-1.4) and (3.31,-0.3) .. (0,0) .. controls (3.31,0.3) and (6.95,1.4) .. (10.93,3.29)   ;

\end{tikzpicture}
    \caption{Different S\&C generated by the 3 different antecedents of $a_8$.}
    \label{fig:s_c_cube_implications}
\end{figure}
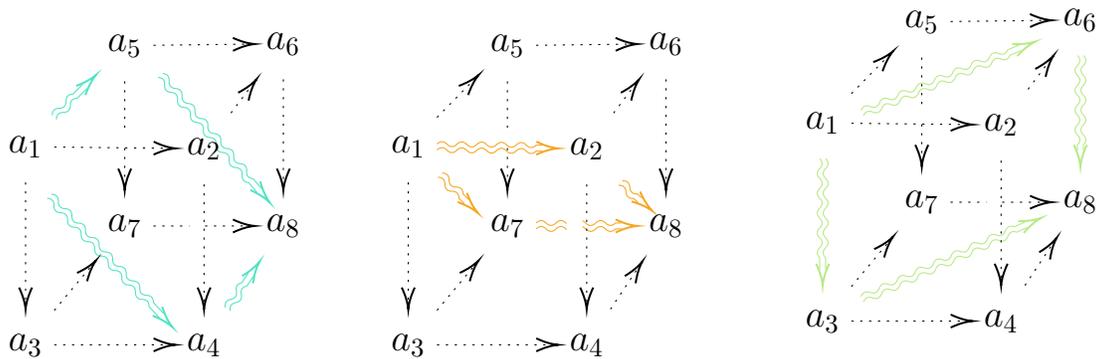

Elements $a_4$, $a_6$ and $a_7$ have 2 antecedents shaping square polytopes, which is analogous to the previous case. However, the last element $a_8$ of this polytope leads to a new situation, as $a_8$ has 3 antecedents ($a_4$, $a_6$ and $a_7$) whose pivots are not antecedents (resp. $a_5$, $a_3$ and $a_2$). Here, each of the 3 antecedents defines a different square polytope with a different fictive element ($a_1$, antecedent, pivot, $\hat{a_8}$), as presented in Figure \ref{fig:s_c_cube_implications}. Can these different S\&C generate different fictive elements? And, if so, can a contrast be defined?

Let us study these three cases. The ``implication principle'' for the contrast in the S\&C means that, for a given antecedent, the relation antecedent/$\hat{a_8}$ is equal to the relation primer/pivot, or, equivalently thanks to commutativity, that the relation primer/antecedent is equal to the relation pivot/$\hat{a_8}$. Here, the three fictive elements are found as:
\begin{itemize}
    \item $\hat{a_8^4} = f_{1/5}.a_4$ in the system $\{a_1, a_4, a_5, a_8\}$.
    \item $\hat{a_8^6} = f_{1/3}.a_6$ in the system $\{a_1, a_2, a_7, a_8\}$.
    \item $\hat{a_8^7} = f_{1/2}.a_7$ in the system $\{a_1, a_3, a_6, a_8\}$.
\end{itemize}

Applying the relation between the primer and the antecedent to the pivot is interesting, because, if the antecedent is itself contrasting in its own square system (for example, $a_4$ in the system $\{a_1, a_2, a_3, a_4\}$), this contrast is also assumed in the relation between the pivot and $\hat{a_8^4}$.

Hence, if several antecedents of $a_8$ are constrasting in their own square systems, different contrasts are assumed to construct the $\hat{a_8^i}$, leading to different fictive elements. There is here an ambiguity on the implication, which needs to be handled.

\paragraph{No antecedent is contrasting} Firstly, let us consider the case where no antecedent of $a_8$ ($a_4, a_6$ and $a_7$) is contrasting. Here, these antecedents are equal to the composition of the relations primer/under-primer of their square systems ($a_4 = f_{1/2}f_{1/3}.a_1, a_6 = f_{1/2}f_{1/5}.a_1$ and $a_7 = f_{1/3}f_{1/5}.a_1$). Hence, the three fictive elements $\hat{a_8^i}$ are all equal to $f_{1/2}f_{1/3}f_{1/5}.a_1$, thanks to commutativity. Figure~\ref{fig:s_c_cube_implications} can help the reader to understand this result.

\paragraph{Only one antecedent is contrasting} Secondly, let us consider the case where \textbf{only one} antecedent is a contrast. In this case, two fictive elements are constructed without contrast as $f_{1/2}f_{1/3}f_{1/5}.a_1$ (as in the precedent case), and the third one replicates the contrast between the primer and this constratic antecedent, as $f_{1/2}f_{1/3}f_{1/5}\gamma.a_1$ (with $\gamma$ denoting the contrastic relation).

Replicating this contrast holds more information than in the non-contrastic cases. Hence, as a rule, the fictive element constructed from the contrastic antecedent is considered as the only valid one.

In both previous cases, a unique valid fictive element $\hat{a_8}$ is constructed to evaluate $a_8$. The equality test $\hat{a_8} == a_8$ then determines the value of $C(a_8 |\{a_1, ..., a_7\}, P)$.

\paragraph{More than one antecedent is contrasting} Finally, when there are at least 2 antecedents with contrasts, it is unclear which fictive element should be chosen. In that case, $a_8$ is considered a constrast. Indeed, if $a_8$ does not admit a valid implication, it cannot be implied, so it is by nature a contrastive element. Hence, $C(a_8 |\{a_1, ..., a_7\}, P) = 1$.

\subsection{General Case, for Regular n-polytopes}
The cube example introduces the case of an element with several antecedents, and the general case extends this principle.

\begin{figure}[ht]
    \centering
\begin{tikzpicture}[x=0.5pt,y=0.5pt,yscale=-1,xscale=1]

\draw    (77.33,132.5) -- (292.59,253.51) ;
\draw [shift={(294.33,254.5)}, rotate = 209.35] [color={rgb, 255:red, 0; green, 0; blue, 0 }  ][line width=0.75]    (10.93,-3.29) .. controls (6.95,-1.4) and (3.31,-0.3) .. (0,0) .. controls (3.31,0.3) and (6.95,1.4) .. (10.93,3.29)   ;
\draw    (247,44.48) -- (456.6,164.5) ;
\draw [shift={(458.33,165.5)}, rotate = 209.8] [color={rgb, 255:red, 0; green, 0; blue, 0 }  ][line width=0.75]    (10.93,-3.29) .. controls (6.95,-1.4) and (3.31,-0.3) .. (0,0) .. controls (3.31,0.3) and (6.95,1.4) .. (10.93,3.29)   ;
\draw    (169.33,131.5) -- (383.57,246.55) ;
\draw [shift={(385.33,247.5)}, rotate = 208.24] [color={rgb, 255:red, 0; green, 0; blue, 0 }  ][line width=0.75]    (10.93,-3.29) .. controls (6.95,-1.4) and (3.31,-0.3) .. (0,0) .. controls (3.31,0.3) and (6.95,1.4) .. (10.93,3.29)   ;
\draw    (74.33,234.5) -- (288.57,349.55) ;
\draw [shift={(290.33,350.5)}, rotate = 208.24] [color={rgb, 255:red, 0; green, 0; blue, 0 }  ][line width=0.75]    (10.93,-3.29) .. controls (6.95,-1.4) and (3.31,-0.3) .. (0,0) .. controls (3.31,0.3) and (6.95,1.4) .. (10.93,3.29)   ;
\draw    (344.33,44.5) -- (558.57,159.55) ;
\draw [shift={(560.33,160.5)}, rotate = 208.24] [color={rgb, 255:red, 0; green, 0; blue, 0 }  ][line width=0.75]    (10.93,-3.29) .. controls (6.95,-1.4) and (3.31,-0.3) .. (0,0) .. controls (3.31,0.3) and (6.95,1.4) .. (10.93,3.29)   ;
\draw    (245.33,143.5) -- (459.57,258.55) ;
\draw [shift={(461.33,259.5)}, rotate = 208.24] [color={rgb, 255:red, 0; green, 0; blue, 0 }  ][line width=0.75]    (10.93,-3.29) .. controls (6.95,-1.4) and (3.31,-0.3) .. (0,0) .. controls (3.31,0.3) and (6.95,1.4) .. (10.93,3.29)   ;
\draw    (167.33,232.5) -- (381.57,347.55) ;
\draw [shift={(383.33,348.5)}, rotate = 208.24] [color={rgb, 255:red, 0; green, 0; blue, 0 }  ][line width=0.75]    (10.93,-3.29) .. controls (6.95,-1.4) and (3.31,-0.3) .. (0,0) .. controls (3.31,0.3) and (6.95,1.4) .. (10.93,3.29)   ;
\draw    (344.33,136.5) -- (558.57,251.55) ;
\draw [shift={(560.33,252.5)}, rotate = 208.24] [color={rgb, 255:red, 0; green, 0; blue, 0 }  ][line width=0.75]    (10.93,-3.29) .. controls (6.95,-1.4) and (3.31,-0.3) .. (0,0) .. controls (3.31,0.3) and (6.95,1.4) .. (10.93,3.29)   ;

\draw (61,117.48) node  [font=\Large]  {$a_{1}$};
\draw (230,31.48) node  [font=\Large]  {$a_{5}$};
\draw (61,217.48) node  [font=\Large]  {$a_{3}$};
\draw (326,31.48) node  [font=\Large]  {$a_{6}$};
\draw (151,118.48) node  [font=\Large]  {$a_{2}$};
\draw (230,125.48) node  [font=\Large]  {$a_{7}$};
\draw (325,126.48) node  [font=\Large]  {$a_{8}$};
\draw (151,217.48) node  [font=\Large]  {$a_{4}$};
\draw (311,255.48) node  [font=\Large]  {$a_{9}$};
\draw (480,169.48) node  [font=\Large]  {$a_{13}$};
\draw (310,355.48) node  [font=\Large]  {$a_{11}$};
\draw (576,168.48) node  [font=\Large]  {$a_{14}$};
\draw (401,256.48) node  [font=\Large]  {$a_{10}$};
\draw (480,263.48) node  [font=\Large]  {$a_{15}$};
\draw (575,264.48) node  [font=\Large]  {$a_{16}$};
\draw (400,355.48) node  [font=\Large]  {$a_{12}$};
\draw    (244.5,31.48) -- (309.5,31.48) ;
\draw [shift={(311.5,31.48)}, rotate = 180] [color={rgb, 255:red, 0; green, 0; blue, 0 }  ][line width=0.75]    (10.93,-3.29) .. controls (6.95,-1.4) and (3.31,-0.3) .. (0,0) .. controls (3.31,0.3) and (6.95,1.4) .. (10.93,3.29)   ;
\draw    (75.5,110.1) -- (213.72,39.76) ;
\draw [shift={(215.5,38.86)}, rotate = 513.03] [color={rgb, 255:red, 0; green, 0; blue, 0 }  ][line width=0.75]    (10.93,-3.29) .. controls (6.95,-1.4) and (3.31,-0.3) .. (0,0) .. controls (3.31,0.3) and (6.95,1.4) .. (10.93,3.29)   ;
\draw    (75.5,117.64) -- (134.5,118.3) ;
\draw [shift={(136.5,118.32)}, rotate = 180.64] [color={rgb, 255:red, 0; green, 0; blue, 0 }  ][line width=0.75]    (10.93,-3.29) .. controls (6.95,-1.4) and (3.31,-0.3) .. (0,0) .. controls (3.31,0.3) and (6.95,1.4) .. (10.93,3.29)   ;
\draw    (165.5,111.27) -- (309.71,39.58) ;
\draw [shift={(311.5,38.69)}, rotate = 513.5699999999999] [color={rgb, 255:red, 0; green, 0; blue, 0 }  ][line width=0.75]    (10.93,-3.29) .. controls (6.95,-1.4) and (3.31,-0.3) .. (0,0) .. controls (3.31,0.3) and (6.95,1.4) .. (10.93,3.29)   ;
\draw    (61,135.48) -- (61,197.48) ;
\draw [shift={(61,199.48)}, rotate = 270] [color={rgb, 255:red, 0; green, 0; blue, 0 }  ][line width=0.75]    (10.93,-3.29) .. controls (6.95,-1.4) and (3.31,-0.3) .. (0,0) .. controls (3.31,0.3) and (6.95,1.4) .. (10.93,3.29)   ;
\draw    (75.5,217.48) -- (134.5,217.48) ;
\draw [shift={(136.5,217.48)}, rotate = 180] [color={rgb, 255:red, 0; green, 0; blue, 0 }  ][line width=0.75]    (10.93,-3.29) .. controls (6.95,-1.4) and (3.31,-0.3) .. (0,0) .. controls (3.31,0.3) and (6.95,1.4) .. (10.93,3.29)   ;
\draw    (165.5,209.9) -- (308.73,134.99) ;
\draw [shift={(310.5,134.06)}, rotate = 512.39] [color={rgb, 255:red, 0; green, 0; blue, 0 }  ][line width=0.75]    (10.93,-3.29) .. controls (6.95,-1.4) and (3.31,-0.3) .. (0,0) .. controls (3.31,0.3) and (6.95,1.4) .. (10.93,3.29)   ;
\draw    (75.5,209.58) -- (213.74,134.33) ;
\draw [shift={(215.5,133.37)}, rotate = 511.44] [color={rgb, 255:red, 0; green, 0; blue, 0 }  ][line width=0.75]    (10.93,-3.29) .. controls (6.95,-1.4) and (3.31,-0.3) .. (0,0) .. controls (3.31,0.3) and (6.95,1.4) .. (10.93,3.29)   ;
\draw    (244.5,125.63) -- (308.5,126.3) ;
\draw [shift={(310.5,126.33)}, rotate = 180.6] [color={rgb, 255:red, 0; green, 0; blue, 0 }  ][line width=0.75]    (10.93,-3.29) .. controls (6.95,-1.4) and (3.31,-0.3) .. (0,0) .. controls (3.31,0.3) and (6.95,1.4) .. (10.93,3.29)   ;
\draw    (230,49.48) -- (230,105.48) ;
\draw [shift={(230,107.48)}, rotate = 270] [color={rgb, 255:red, 0; green, 0; blue, 0 }  ][line width=0.75]    (10.93,-3.29) .. controls (6.95,-1.4) and (3.31,-0.3) .. (0,0) .. controls (3.31,0.3) and (6.95,1.4) .. (10.93,3.29)   ;
\draw    (325.81,49.48) -- (325.21,106.48) ;
\draw [shift={(325.19,108.48)}, rotate = 270.6] [color={rgb, 255:red, 0; green, 0; blue, 0 }  ][line width=0.75]    (10.93,-3.29) .. controls (6.95,-1.4) and (3.31,-0.3) .. (0,0) .. controls (3.31,0.3) and (6.95,1.4) .. (10.93,3.29)   ;
\draw    (151,136.48) -- (151,197.48) ;
\draw [shift={(151,199.48)}, rotate = 270] [color={rgb, 255:red, 0; green, 0; blue, 0 }  ][line width=0.75]    (10.93,-3.29) .. controls (6.95,-1.4) and (3.31,-0.3) .. (0,0) .. controls (3.31,0.3) and (6.95,1.4) .. (10.93,3.29)   ;
\draw    (498.5,169.29) -- (555.5,168.69) ;
\draw [shift={(557.5,168.67)}, rotate = 539.4] [color={rgb, 255:red, 0; green, 0; blue, 0 }  ][line width=0.75]    (10.93,-3.29) .. controls (6.95,-1.4) and (3.31,-0.3) .. (0,0) .. controls (3.31,0.3) and (6.95,1.4) .. (10.93,3.29)   ;
\draw    (325.5,248.1) -- (459.72,179.8) ;
\draw [shift={(461.5,178.89)}, rotate = 513.03] [color={rgb, 255:red, 0; green, 0; blue, 0 }  ][line width=0.75]    (10.93,-3.29) .. controls (6.95,-1.4) and (3.31,-0.3) .. (0,0) .. controls (3.31,0.3) and (6.95,1.4) .. (10.93,3.29)   ;
\draw    (325.5,255.64) -- (380.5,256.25) ;
\draw [shift={(382.5,256.27)}, rotate = 180.64] [color={rgb, 255:red, 0; green, 0; blue, 0 }  ][line width=0.75]    (10.93,-3.29) .. controls (6.95,-1.4) and (3.31,-0.3) .. (0,0) .. controls (3.31,0.3) and (6.95,1.4) .. (10.93,3.29)   ;
\draw    (419.5,247.18) -- (555.71,178.68) ;
\draw [shift={(557.5,177.78)}, rotate = 513.3] [color={rgb, 255:red, 0; green, 0; blue, 0 }  ][line width=0.75]    (10.93,-3.29) .. controls (6.95,-1.4) and (3.31,-0.3) .. (0,0) .. controls (3.31,0.3) and (6.95,1.4) .. (10.93,3.29)   ;
\draw    (310.82,273.48) -- (310.2,335.48) ;
\draw [shift={(310.18,337.48)}, rotate = 270.57] [color={rgb, 255:red, 0; green, 0; blue, 0 }  ][line width=0.75]    (10.93,-3.29) .. controls (6.95,-1.4) and (3.31,-0.3) .. (0,0) .. controls (3.31,0.3) and (6.95,1.4) .. (10.93,3.29)   ;
\draw    (328.5,355.48) -- (379.5,355.48) ;
\draw [shift={(381.5,355.48)}, rotate = 180] [color={rgb, 255:red, 0; green, 0; blue, 0 }  ][line width=0.75]    (10.93,-3.29) .. controls (6.95,-1.4) and (3.31,-0.3) .. (0,0) .. controls (3.31,0.3) and (6.95,1.4) .. (10.93,3.29)   ;
\draw    (418.5,345.86) -- (554.73,275.02) ;
\draw [shift={(556.5,274.1)}, rotate = 512.53] [color={rgb, 255:red, 0; green, 0; blue, 0 }  ][line width=0.75]    (10.93,-3.29) .. controls (6.95,-1.4) and (3.31,-0.3) .. (0,0) .. controls (3.31,0.3) and (6.95,1.4) .. (10.93,3.29)   ;
\draw    (328.5,345.47) -- (459.74,274.44) ;
\draw [shift={(461.5,273.49)}, rotate = 511.58] [color={rgb, 255:red, 0; green, 0; blue, 0 }  ][line width=0.75]    (10.93,-3.29) .. controls (6.95,-1.4) and (3.31,-0.3) .. (0,0) .. controls (3.31,0.3) and (6.95,1.4) .. (10.93,3.29)   ;
\draw    (498.5,263.67) -- (554.5,264.26) ;
\draw [shift={(556.5,264.28)}, rotate = 180.6] [color={rgb, 255:red, 0; green, 0; blue, 0 }  ][line width=0.75]    (10.93,-3.29) .. controls (6.95,-1.4) and (3.31,-0.3) .. (0,0) .. controls (3.31,0.3) and (6.95,1.4) .. (10.93,3.29)   ;
\draw    (480,187.48) -- (480,243.48) ;
\draw [shift={(480,245.48)}, rotate = 270] [color={rgb, 255:red, 0; green, 0; blue, 0 }  ][line width=0.75]    (10.93,-3.29) .. controls (6.95,-1.4) and (3.31,-0.3) .. (0,0) .. controls (3.31,0.3) and (6.95,1.4) .. (10.93,3.29)   ;
\draw    (575.81,186.48) -- (575.21,244.48) ;
\draw [shift={(575.19,246.48)}, rotate = 270.6] [color={rgb, 255:red, 0; green, 0; blue, 0 }  ][line width=0.75]    (10.93,-3.29) .. controls (6.95,-1.4) and (3.31,-0.3) .. (0,0) .. controls (3.31,0.3) and (6.95,1.4) .. (10.93,3.29)   ;
\draw    (400.82,274.48) -- (400.2,335.48) ;
\draw [shift={(400.18,337.48)}, rotate = 270.58] [color={rgb, 255:red, 0; green, 0; blue, 0 }  ][line width=0.75]    (10.93,-3.29) .. controls (6.95,-1.4) and (3.31,-0.3) .. (0,0) .. controls (3.31,0.3) and (6.95,1.4) .. (10.93,3.29)   ;

\end{tikzpicture}

    \caption{4-polytope}
    \label{fig:4_polytope}
\end{figure}
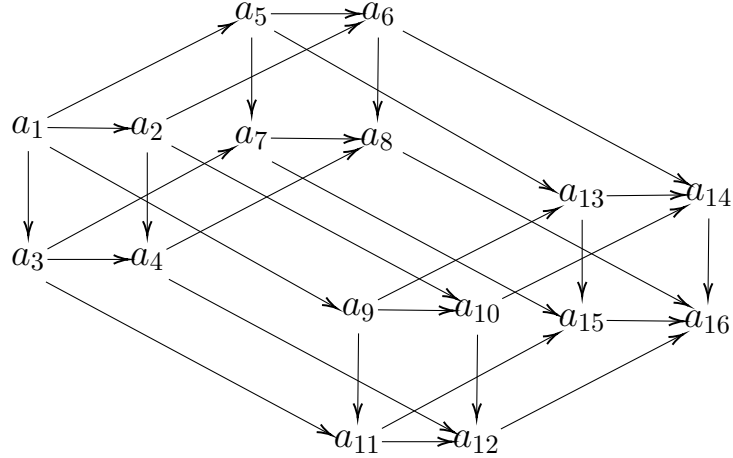

To simplify visualizations, let us consider the 4-polytope case, represented in Figure~\ref{fig:4_polytope}. In this polytope, $a_8$ is no longer the only element with several couples antecedent/pivot, and is itself an antecedent of $a_{16}$.

When $a_8$ admits a unique valid fictive element $\hat{a_8}$ (only 0 or 1 contrast among its antecedents, as seen previously), the previous case can be extended by checking if $a_8$ is itself a contrast or not ($\hat{a_8} == a_8$).

When $a_8$ does not admit a fictive element, it is a contrast. In both cases, $\hat{a_{16}^8}$ can be constructed as $f_{1/8}.a_9$, and, counting how many constrastive antecedents $a_{16}$ holds, the evaluation process for $a_8$, presented above, can be extended (with 4 antecedents instead of 3).

Hence, for any element $a_i$, if 0 or 1 of its antecedents is a contrast, a unique valid fictive element can be constructed, and $C(a_i |\{a_1, ..., a_{i-1}\}, P)$ depends on the relation $f_{\hat{i}/i}$. Otherwise, when several antecedents are contrasting, fictive elements are ambiguous, thus $a_i$ is treated itself as a contrast, both for the cost $C(a_i |\{a_1, ..., a_{i-1}\}, P)$ and for its successors.

\begin{definition}[Valid antecedents (set)] We call ``valid antecedents'' the set of all the antecedents of an element which can be used to construct a fictive element without ambiguity. Hence, this set can contain all antecedents of an element (if none of them is a contrast), only one (if it is the only contrast) or be empty. It is denoted $V_i$.
\end{definition}


The key point is to update the set of valid antecedents when facing a contrastic element: only this contrastic element has to be considered as valid for its successors. Concretely, this is made by intersecting each set of valid antecedents of its successors with this element (and its pivot). This indeed results in an empty set when several antecedents are contrastive.

The aforementioned process results in the general complexity cost function for regular polytopes, presented in Algorithm~\ref{alg:implication_principle}. In addition to previous definitions, let us denote $S_i$ the set of successors of an element $a_i$.

\begin{algorithm}[ht]
\caption{Implication principle for an element $a_i$.}
\begin{algorithmic}
\algnewcommand\algorithmicinput{\textbf{Input:}}
\algnewcommand\Input{\item[\algorithmicinput]}
\algnewcommand\algorithmicoutput{\textbf{Output:}}
\algnewcommand\Output{\item[\algorithmicoutput]}
\Input $a_i$, element of a polytope, with a set of valid antecedents $V_i$, and relations $f \in G_r$, the current cost of the polytope $c$.
\Output Updated cost $c$

\If {$\exists a_j \in V_i / f(a_1, a_j) = f(p_i^j, a_i)$}
\State \Comment{Looking at its valid antecedents, and searching for an implication without contrast.}
            \State $c = c$ \Comment{Information is redundant.}

\Else \Comment{There is no implication without contrast (including the case where $V_i$ is empty)}
    \State $c = c + 1$ \Comment{This element needs to be encoded.}
    \For{$s \in S_i$} \Comment{Iterating over the successors of $a_i$}
        \State $V_s = V_s \cap \{a_i, p_s^i\}$
        \State \Comment{Updating the antecedents of this successor, to keep only $a_i$ and its pivot.}
    \EndFor
\EndIf

\Return c
\end{algorithmic}
\label{alg:implication_principle}
\end{algorithm}

\subsection{Irregular Polytopes}
Finally, the complexity cost function can be extended to any irregular polytope. Starting with a n-polytope, an irregular polytope is constructed by deleting and/or adding another regular polytope of smaller dimension which contains the last element of the n-polytope.

As stated before, this condition ensures that, when deleting an element, all of its successors are also deleted. This also ensures that deletion does not break the previously designed rule. Thus, deletion only reduces the number of successors of some elements, but does not change the aforementioned rule.

\begin{figure}[ht!]
  \begin{center}
\begin{tikzpicture}[x=0.75pt,y=0.75pt,yscale=-1,xscale=1]

\draw [color={rgb, 255:red, 8; green, 57; blue, 139 }  ,draw opacity=1 ][line width=2.25]  [dash pattern={on 6.75pt off 4.5pt}]  (281.58,237.45) -- (296.57,256.1)(279.24,239.33) -- (294.23,257.98) ;
\draw [shift={(301.67,264.83)}, rotate = 231.2] [color={rgb, 255:red, 8; green, 57; blue, 139 }  ,draw opacity=1 ][line width=2.25]    (17.49,-5.26) .. controls (11.12,-2.23) and (5.29,-0.48) .. (0,0) .. controls (5.29,0.48) and (11.12,2.23) .. (17.49,5.26)   ;
\draw [color={rgb, 255:red, 32; green, 148; blue, 131 }  ,draw opacity=1 ][line width=2.25]  [dash pattern={on 2.53pt off 3.02pt}]  (322.67,265.83) -- (345.89,241.71) ;
\draw [shift={(348.67,238.83)}, rotate = 493.92] [color={rgb, 255:red, 32; green, 148; blue, 131 }  ,draw opacity=1 ][line width=2.25]    (17.49,-5.26) .. controls (11.12,-2.23) and (5.29,-0.48) .. (0,0) .. controls (5.29,0.48) and (11.12,2.23) .. (17.49,5.26)   ;
\draw [color={rgb, 255:red, 8; green, 57; blue, 139 }  ,draw opacity=1 ][line width=2.25]  [dash pattern={on 6.75pt off 4.5pt}]  (322.8,190.85) -- (336.68,206.72)(320.54,192.82) -- (334.42,208.69) ;
\draw [shift={(342.13,215.23)}, rotate = 228.82999999999998] [color={rgb, 255:red, 8; green, 57; blue, 139 }  ,draw opacity=1 ][line width=2.25]    (17.49,-5.26) .. controls (11.12,-2.23) and (5.29,-0.48) .. (0,0) .. controls (5.29,0.48) and (11.12,2.23) .. (17.49,5.26)   ;

\draw (183.07,128.51) node  [font=\Large]  {$a_{1}$};
\draw (234.12,81.34) node  [font=\Large]  {$a_{6}$};
\draw (183.07,218.27) node  [font=\Large]  {$a_{3}$};
\draw (310.48,81.8) node  [font=\Large]  {$a_{7}$};
\draw (262.4,129.34) node  [font=\Large]  {$a_{2}$};
\draw (235.28,176.97) node  [font=\Large]  {$a_{8}$};
\draw (308.95,176.63) node  [font=\Large]  {$a_{9}$};
\draw (261.4,218.27) node  [font=\Large]  {$a_{4}$};
\draw (360.53,221.74) node  [font=\Large]  {$a_{10}$};
\draw (313.2,276.34) node  [font=\Large]  {$a_{5}$};
\draw [color={rgb, 255:red, 230; green, 172; blue, 90 }  ,draw opacity=1 ][line width=1.5]    (249.12,81.43) -- (292.48,81.69) ;
\draw [shift={(295.48,81.71)}, rotate = 180.35] [color={rgb, 255:red, 230; green, 172; blue, 90 }  ,draw opacity=1 ][line width=1.5]    (14.21,-4.28) .. controls (9.04,-1.82) and (4.3,-0.39) .. (0,0) .. controls (4.3,0.39) and (9.04,1.82) .. (14.21,4.28)   ;
\draw [color={rgb, 255:red, 32; green, 148; blue, 131 }  ,draw opacity=1 ][line width=2.25]  [dash pattern={on 2.53pt off 3.02pt}]  (198.07,114.65) -- (216.19,97.91) ;
\draw [shift={(219.12,95.2)}, rotate = 497.26] [color={rgb, 255:red, 32; green, 148; blue, 131 }  ,draw opacity=1 ][line width=2.25]    (17.49,-5.26) .. controls (11.12,-2.23) and (5.29,-0.48) .. (0,0) .. controls (5.29,0.48) and (11.12,2.23) .. (17.49,5.26)   ;
\draw [color={rgb, 255:red, 230; green, 172; blue, 90 }  ,draw opacity=1 ][line width=1.5]    (198.07,128.67) -- (244.4,129.15) ;
\draw [shift={(247.4,129.18)}, rotate = 180.6] [color={rgb, 255:red, 230; green, 172; blue, 90 }  ,draw opacity=1 ][line width=1.5]    (14.21,-4.28) .. controls (9.04,-1.82) and (4.3,-0.39) .. (0,0) .. controls (4.3,0.39) and (9.04,1.82) .. (14.21,4.28)   ;
\draw [color={rgb, 255:red, 32; green, 148; blue, 131 }  ,draw opacity=1 ][line width=2.25]  [dash pattern={on 2.53pt off 3.02pt}]  (277.4,114.51) -- (292.64,99.45) ;
\draw [shift={(295.48,96.63)}, rotate = 495.33] [color={rgb, 255:red, 32; green, 148; blue, 131 }  ,draw opacity=1 ][line width=2.25]    (17.49,-5.26) .. controls (11.12,-2.23) and (5.29,-0.48) .. (0,0) .. controls (5.29,0.48) and (11.12,2.23) .. (17.49,5.26)   ;
\draw [color={rgb, 255:red, 213; green, 65; blue, 90 }  ,draw opacity=1 ][line width=1.5]    (184.57,146.51) -- (184.57,191.27)(181.57,146.51) -- (181.57,191.27) ;
\draw [shift={(183.07,200.27)}, rotate = 270] [color={rgb, 255:red, 213; green, 65; blue, 90 }  ,draw opacity=1 ][line width=1.5]    (14.21,-4.28) .. controls (9.04,-1.82) and (4.3,-0.39) .. (0,0) .. controls (4.3,0.39) and (9.04,1.82) .. (14.21,4.28)   ;
\draw [color={rgb, 255:red, 230; green, 172; blue, 90 }  ,draw opacity=1 ][line width=1.5]    (198.07,218.27) -- (243.4,218.27) ;
\draw [shift={(246.4,218.27)}, rotate = 180] [color={rgb, 255:red, 230; green, 172; blue, 90 }  ,draw opacity=1 ][line width=1.5]    (14.21,-4.28) .. controls (9.04,-1.82) and (4.3,-0.39) .. (0,0) .. controls (4.3,0.39) and (9.04,1.82) .. (14.21,4.28)   ;
\draw [color={rgb, 255:red, 32; green, 148; blue, 131 }  ,draw opacity=1 ][line width=2.25]  [dash pattern={on 2.53pt off 3.02pt}]  (276.4,205.14) -- (290.95,192.4) ;
\draw [shift={(293.95,189.77)}, rotate = 498.79] [color={rgb, 255:red, 32; green, 148; blue, 131 }  ,draw opacity=1 ][line width=2.25]    (17.49,-5.26) .. controls (11.12,-2.23) and (5.29,-0.48) .. (0,0) .. controls (5.29,0.48) and (11.12,2.23) .. (17.49,5.26)   ;
\draw [color={rgb, 255:red, 32; green, 148; blue, 131 }  ,draw opacity=1 ][line width=2.25]  [dash pattern={on 2.53pt off 3.02pt}]  (198.07,206.41) -- (217.14,191.32) ;
\draw [shift={(220.28,188.84)}, rotate = 501.65] [color={rgb, 255:red, 32; green, 148; blue, 131 }  ,draw opacity=1 ][line width=2.25]    (17.49,-5.26) .. controls (11.12,-2.23) and (5.29,-0.48) .. (0,0) .. controls (5.29,0.48) and (11.12,2.23) .. (17.49,5.26)   ;
\draw [color={rgb, 255:red, 230; green, 172; blue, 90 }  ,draw opacity=1 ][line width=1.5]    (250.28,176.9) -- (290.95,176.71) ;
\draw [shift={(293.95,176.7)}, rotate = 539.73] [color={rgb, 255:red, 230; green, 172; blue, 90 }  ,draw opacity=1 ][line width=1.5]    (14.21,-4.28) .. controls (9.04,-1.82) and (4.3,-0.39) .. (0,0) .. controls (4.3,0.39) and (9.04,1.82) .. (14.21,4.28)   ;
\draw [color={rgb, 255:red, 213; green, 65; blue, 90 }  ,draw opacity=1 ][line width=1.5]    (235.84,99.32) -- (236.45,149.96)(232.84,99.35) -- (233.45,149.99) ;
\draw [shift={(235.06,158.97)}, rotate = 269.31] [color={rgb, 255:red, 213; green, 65; blue, 90 }  ,draw opacity=1 ][line width=1.5]    (14.21,-4.28) .. controls (9.04,-1.82) and (4.3,-0.39) .. (0,0) .. controls (4.3,0.39) and (9.04,1.82) .. (14.21,4.28)   ;
\draw [color={rgb, 255:red, 213; green, 65; blue, 90 }  ,draw opacity=1 ][line width=1.5]    (311.69,99.83) -- (310.89,149.66)(308.69,99.78) -- (307.89,149.61) ;
\draw [shift={(309.24,158.63)}, rotate = 270.92] [color={rgb, 255:red, 213; green, 65; blue, 90 }  ,draw opacity=1 ][line width=1.5]    (14.21,-4.28) .. controls (9.04,-1.82) and (4.3,-0.39) .. (0,0) .. controls (4.3,0.39) and (9.04,1.82) .. (14.21,4.28)   ;
\draw [color={rgb, 255:red, 213; green, 65; blue, 90 }  ,draw opacity=1 ][line width=1.5]    (263.7,147.36) -- (263.2,191.29)(260.7,147.32) -- (260.2,191.26) ;
\draw [shift={(261.6,200.27)}, rotate = 270.64] [color={rgb, 255:red, 213; green, 65; blue, 90 }  ,draw opacity=1 ][line width=1.5]    (14.21,-4.28) .. controls (9.04,-1.82) and (4.3,-0.39) .. (0,0) .. controls (4.3,0.39) and (9.04,1.82) .. (14.21,4.28)   ;

\end{tikzpicture}

\end{center}
  \caption{Visualization for addition}
  \label{fig:irregular_addition}

\end{figure}
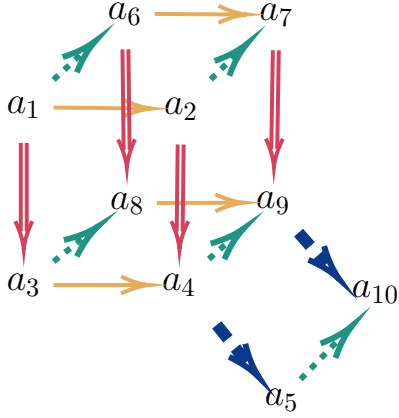

Nonetheless, addition in a polytope adds a new case to the general rule, presented with help of Figure~\ref{fig:irregular_addition}.

When an element is added to the polytope, the element on which it is attached is its antecedent (here for instance, $a_4$ is an antecedent of $a_5$, and equivalently for $a_9$ and $a_{10}$). However, these antecedence relations do not define a pivot element, as relations $f_{4/5}$ and $f_{9/10}$ do not have a parallel relation starting on the primer.

In that case, relations are compared with the identity function for the complexity cost. In practice, it can be obtained by considering that the pivot of $a_5$ related to its antecedent $a_4$ is the primer $a_1$, which follows the general rule.

Additionally, $a_5$ is also antecedent of $a_{10}$, and relation $f_{5/10}$ has a parallel relation $f_{1/6}$ starting from the primer, so $a_6$ is a pivot for $a_{10}$ related to $a_5$, which follows the general rule.

\subsection{Computing the Cost of a Polytope}
Finally, Algorithm~\ref{alg:final_algo} presents the general complexity cost algorithm for a sequence of musical elements on a polytope.


\begin{algorithm}[ht]
\caption{Compression cost of a polytope.}
\begin{algorithmic}
\algnewcommand\algorithmicinput{\textbf{Input:}}
\algnewcommand\Input{\item[\algorithmicinput]}
\algnewcommand\algorithmicoutput{\textbf{Output:}}
\algnewcommand\Output{\item[\algorithmicoutput]}
\Input Polytope with vertices $\{a_i\} \in A, i = {1, ..., m}$, and an abelian group $G_r$.
\Output Cost $c$
\State c = 1 \Comment{Cost to encode $a_1$}
\For{$i = 2, ..., m$}
\State $V_i = \{$antecedents for $a_i\}$ \Comment{Initializing valid antecedents for all elements with all their antecedents}
\EndFor

\For{$i = 2, ..., m$} \Comment{Iterating over the elements of the polytope}

    \If{$V_i = \{a_1\}$} \Comment{If this element is an under-primer}
        \If{$a_1 = a_i$}
            \State $c = c$ \Comment{Information is redundant.}
        \Else
            \State $c = c + 1$ \Comment{This element needs to be encoded.}
        \EndIf

    \Else \Comment{This element is not an under-primer}

        \If {$\exists a_j \in V_i / f(a_1, a_j) = f(p_i^j, a_i)$}
            \State \Comment{Looking at its antecedents, and searching for an implication without contrast.}
            \State $c = c$ \Comment{Information is redundant.}

        \Else \Comment{There is no implication without contrast (including the case where $V_i$ is empty)}
            \State $c = c + 1$ \Comment{This element needs to be encoded.}
            \For{$s \in S_i$} \Comment{Iterating over the successors of $a_i$}
                \State $V_s = V_s \cap \{a_i, p_s^i\}$
                \State \Comment{Updating the antecedents of this successor, to keep only $a_i$ and its pivot.}
            \EndFor
        \EndIf

    \EndIf

\EndFor
\Return c
\end{algorithmic}
\label{alg:final_algo}
\end{algorithm}

\section{Numerical Experiments}
\label{sec:experiences}
Algorithm~\ref{alg:final_algo} is developed in Python, along with a model handling polytopes, and is open-source\footnote{https://gitlab.inria.fr/amarmore/musiconpolytopes}~\cite{PolytopeToolbox}. Results are based on the RWC Pop database~\cite{goto2002rwc}.

\subsection{Data}
This algorithm has been tested in a same manner than C. Guichaoua in his PhD thesis~\cite{theseGuichaoua}, which introduces the polytopic analysis of music. Particularly, tests are conducted on the semiotic database of annotation for RWC Pop\footnote{which can be found at https://gitlab.inria.fr/amarmore/rwc\_quaero\_annotations}~\cite[Chap.3.3]{theseGuichaoua}. This database contains beatwise aligned chord annotations, obtained from the initial annotations of the RWC Pop database~\cite{goto2002rwc} (``auto''), which were then manually corrected and homogenized by a human annotator (``manual'').

In these annotations, each song is represented by a discretized sequence of perfect chords, synchronized on beats of the song. As a first attempt, silences were replaced with the previous chord (or the first chord of the song if silences are opening it). Defining a relation between a chord and a silence could be explored in future work.

\subsection{Penalties}
Section~\ref{sec:guichaoua} presents the raw polytopic cost $C(S,P)$ for a musical passage $S$ on a polytope $P$. Two penalty costs are added to this raw polytopic cost such that:
\begin{equation}
    \mathscr{C}(S) = \min_{P} \left(C(S,P) + f_a(P)\right) + f_r(S)
\end{equation}
iterating over all polytopes $P$ containing $card(S)$ vertices, size of the musical passage.

\subsubsection{Alteration Penalty $f_a$}
A first penalty is applied to the polytope itself, related to its irregularities. In an information theory-like viewpoint, as presented in Section~\ref{sec:guichaoua}, a polytope can be defined by a quantity of information. Regular polytopes can be entirely described by their dimension, while altering a regular polytope (either by deletion or addition) requires to encode the shape and the position of the alteration.

In that sense, altering a polytope $P$ increases the complexity. This increase is handled by adding a penalty $f_a(P) = p_a \in \mathbb{R}_+$ to the raw score when the polytope is irregular by either an addition or a deletion, and by adding $f_a(P) = 2 p_a$ when the polytope is altered by both addition and deletion. Parameter $p_a$ is fitted in experiments.

\subsubsection{Regularity Penalty $f_r$}
The second penalty considers the size of the segment. Indeed, as presented in~\cite{sargent2016estimating}, some segment sizes are more frequent than other in the RWC Pop database, particularly segments of 32 beats. Sargent et al. shows that adding a penalty prior in segmentation algorithms can enhance segmentation scores. For consistency with~\cite{theseGuichaoua}, we use the function $f_r (S) = p_r |card(S) - 32|$ as a regularization for the segment $S$ of size $card(S)$. Parameter $p_r \in \mathbb{R}_+$ is fitted in experiments.

\subsection{Scores}
The goal of the task is to retrieve frontiers between structural segments, \textit{i.e.} beats on which the segment is changing.

These estimated frontiers are compared with the annotation in order to compute True Positive, False Positive (wrong estimation of a frontier) and False Negative (frontier not found in estimation) rates. From this rates are computed Precision, Recall and F1-measure, as presented in~\cite{4432648}. A frontier is considered correct if it is exact or falls close enough (within a tolerance window) to an annotated frontier. These experiments were restricted to 0 and 3 beats tolerance windows, as in~\cite{theseGuichaoua}.

\begin{table}[]
\centering
\begin{tabular}{|l|l|l|l|l|l|l|l|l|}
\hline
Technique                                         &                      & $P_0$  & $R_0$  & $F_0$  & $P_3$  & $R_3$  & $F_3$  & Computation time \\ \hline
\multirow{2}{*}{MusicOnPolytopes~\cite{PolytopeToolbox}}                    & $p_a = 0, p_r = 0$   & 50.3\% & 61.8\% & 55.1\% & 55\%   & 68\%   & 60.4\% & 3 $\frac{1}{2}$ hours     \\ \cline{2-9}
                                                  & $p_a = 3, p_r = 0.1$ & 68.2\% & 73.6\% & 70.6\% & 68.9\% & 74.5\% & 71.4\% & 3 $\frac{1}{2}$ hours     \\ \hline
\multirow{2}{*}{Results from~\cite{theseGuichaoua}}                & $p_a = 0, p_r = 0$   &   -  & -  & 43.7\% &  -      &  -      &  -      & Not mentionned   \\ \cline{2-9}
                                                  & Optimal conditions*  &    -    &    -    & 69\%   &    -    &     -   & 70\%   & Not mentionned   \\ \hline
Code of~\cite{theseGuichaoua}, & $p_a = 0, p_r = 0$   & 35.8\% & 56.7\% & 43.3\% & 39.6\% & 62.9\% & 47.9\% & 8 hours          \\ \cline{2-9}
 on author's laptop     & Optimal conditions*  & 59.2\% & 63.4\% & 61.1\% & 61.2\% & 65.6\% & 63.2\% & 8 hours          \\ \hline
\end{tabular}
\caption{Numerical experiences on database ``Manual''. *Optimal conditions refer to the optimal conditions of~\cite{theseGuichaoua}, which are slightly different than in our model. C. Guichaoua indeed considered that $p_a$ should be different when considering addition and deletion, leading to two parameters $p_a^+$ and $p_a^-$, and also that $p_r$ should distinguish sizes larger and lower than 32, leading to two parameters $p_r^+$ and $p_r^-$. These optimal conditions hence refer to $p_a^+ = 2.25, p_a^- = 3, p_r^+ = 0, p_r^- = 0.125$.}

\label{tab:results_guichaoua_manual}
\end{table}

\begin{table}[]
\begin{tabular}{|l|l|l|l|l|l|l|l|}
\hline
Technique                                         &                     & $P_0$ & $R_0$  & $F_0$  & $P_3$  & $R_3$  & $F_3$   \\ \hline
\multirow{2}{*}{MusicOnPolytopes~\cite{PolytopeToolbox}}                    & $p_a = 0, p_r = 0$  & 29\%  & 39.5\% & 33.1\% & 42.6\% & 59.8\% & 49.2\% \\ \cline{2-8}
 & $p_a = 4, p_r = 0.2$ & 44.5\% & 47.1\% & 45.6\%&56.2\%&60\%&57.8\% \\ \hline
\multirow{2}{*}{Results from~\cite{theseGuichaoua}}                & $p_a = 0, p_r = 0$  &   -    &     -   &    -    & - &   -     &     -    \\ \cline{2-8}
                                                  & Optimal conditions* &     -  &     -   & 37.4\% &      -  &   -     & 55.2\%   \\ \hline
Code of~\cite{theseGuichaoua}, & $p_a = 0, p_r = 0$  &23.6\% & 39\% & 28.9\% &35\% & 59.5\% & 43.4\%  \\ \cline{2-8}
on author's laptop & Optimal conditions* &41.7\% &44.4\% & 42.8\% &53.9\% & 57.7\% & 55.5\% \\ \hline
\end{tabular}
\caption{Numerical experiments on the ``Auto'' database. *Optimal conditions refer to the optimal conditions of~\cite{theseGuichaoua}, equal to: $p_a^+ = 2.5, p_a^- = 2.5, p_r^+ = 0, p_r^- = 0.125$.}
\label{tab:results_guichaoua_auto}

\end{table}

\begin{table}[]

\begin{tabular}{|l|l|l|l|l|l|l|l|l|}
\hline
Database                & \multicolumn{2}{l|}{Technique}                         & $P_0$  & $R_0$  & $F_0$  & $P_3$  & $R_3$  & $F_3$  \\ \hline \hline
\multirow{4}{*}{Manual} & \multirow{2}{*}{Triad circle} & $p_a = 0, p_r = 0$     & 50.3\% & 61.8\% & 55.1\% & 55\%   & 68\%   & 60.4\% \\ \cline{3-9}
                        &                               & $p_a = 3, p_r = 0.1$   & 68.2\% & 73.6\% & 70.6\% & 68.9\% & 74.5\% & 71.4\% \\ \cline{2-9}
                        & \multirow{2}{*}{Tonnetz}      & $p_a = 0, p_r = 0$     & 50\%   & 61.2\% & 54.7\% & 55.3\% & 68.4\% & 60.7\% \\ \cline{3-9}
                        &                               & $p_a = 3.5, p_r = 0.1$ & 67.1\% & 72.5\% & 69.5\% & 68\%   & 73.6\% & 70.5\% \\ \hline \hline
\multirow{4}{*}{Auto}   & \multirow{2}{*}{Triad circle} & $p_a = 0, p_r = 0$     & 29\%   & 39.5\% & 33.1\% & 42.6\% & 59.8\% & 49.2\% \\ \cline{3-9}
                        &                               & $p_a = 4, p_r = 0.2$   & 44.5\% & 47.1\% & 45.6\% & 56.2\% & 60\%   & 57.8\% \\ \cline{2-9}
                        & \multirow{2}{*}{Tonnetz}      & $p_a = 0, p_r = 0$     & 27.1\% & 36.6\% & 30.8\% & 41.1\% & 57.7\% & 47.5\% \\ \cline{3-9}
                        &                               & $p_a = 4, p_r = 0.3$   & 44.8\% & 46.3\% & 45.4\% & 56.5\% & 58.7\% & 57.4\% \\ \hline
\end{tabular}
\caption{Results of MusicOnPolytopes~\cite{PolytopeToolbox}. Comparison between triad circle and tonnetz relations.}

\label{tab:results_tonnetz_triad}

\end{table}
Results presented in tables \ref{tab:results_guichaoua_manual} and \ref{tab:results_guichaoua_auto} are computed using the triad circle model of relations, which is common to both works. Results obtained with the new \textit{MusicOnPolytopes} toolbox are higher than those of~\cite{theseGuichaoua}. At this time, the differences are difficult to explain.

In addition, when running the code of~\cite{theseGuichaoua} (obtained from C. Guichaoua himself), the results on the Manual database are worst than the ones presented in~\cite{theseGuichaoua}. These results may be due to downgrading or modifications of external libraries since its initial development, but are also hard to explain.

In addition, Table~\ref{tab:results_tonnetz_triad} compares segmentation results obtained either with the triad circle or the tonnetz as relation model. Results are not significantly different between the two models of relations, but the tonnetz obtains generally worst results than the triad circle.

\section{Conclusion}
In conclusion, this article presents a new code framework, in Python, to compute polytopic analysis of music, introduced in~\cite{theseGuichaoua}. This framework shows interesting results when applied on the structural segmentation task of symbolic music.

This work shows an improvement in performance compared to those obtained by C. Guichaoua in~\cite{theseGuichaoua}, and calls for further development. In particular, an exciting lead would be the development of a new group of relation $G_r$ for discretized audio signals, in order to extend this work for the structural segmentation of audio signals.

\bibliographystyle{ieeetr}
\bibliography{references}

\appendix
\section{Computational Representation of Polytopes}
A polytope can be represented either by its geometrical model (n-hypercube with alteration), or by a symbolic representation, more suited for computational treatment.

In his thesis~\cite{theseGuichaoua}, C. Guichaoua uses binary trees to model polytopes, where leafs represent final elements.

In MusicOnPolytopes, polytopes are represented by nested lists. An element of a polytope is represented by a ``1'', and every dimension represents a level of nesting of this element. For instance, a 1-polytope, linking two elements, is represented as [1,1], and a 2-polytope, with 4 elements, is represented as [[1,1], [1,1]].

For irregularities, a deletion of an element is represented by the deletion of a ``1'' in this list, for instance the 2-polytope with the 0-polytope deletion is represented as [[1,1], [1]]. An addition is represented by a tuple, signifying on which vertex is attached the new element, for instance the 1-polytope with the 0-polytope addition is represented as [1, (1,1)].

These general polytopes can then be adapted to a particular musical sequence (for instance [[Ab, Ab],[Gm,Gm]]), or extended to indexed polytope, where each element represents   the index of the element in the polytope (for instance [[0,1], [2,3]]).

A tutorial notebook is present with the code\footnote{https://gitlab.inria.fr/amarmore/musiconpolytopes/-/blob/master/Notebooks/Tutorial - Handling polytopes.ipynb}.

\end{document}